**GLASSES IN COLLOIDAL SYSTEMS.**

**ATTRACTIVE INTERACTIONS AND GELATION.**

*Antonio M. Puertas*

*Group of Complex Fluids Physics, Departamento de Física Aplicada, Universidad de Almería.*

*04120 Almería, Andalucía, SPAIN*

*Matthias Fuchs*

*Fachbereich Physik, Universität Konstanz, D-78457 Konstanz, GERMANY*

*ABSTRACT*

In this chapter, a study of the glass transitions in colloidal systems is presented, in connection with gelation, mainly from theoretical and simulation results. Mode Coupling Theory, which anticipated the existence of attraction driven glasses, is reviewed, and its predictions concerning attractive glasses discussed. Results from computer simulations will be presented for different models and the predictions of the theory will be tested. Starting from high density, where reasonable agreement is found, the study will be extended to low density, where new modes for the decay of density correlation functions appear. In all cases, the results will be also be brought into connection with experiments, and we will conclude with a discussion of the present understanding of the mechanisms leading to gelation.

## 1. INTRODUCTION

When an atomic systems is cooled below its glass temperature, it vitrifies, i.e. forms an amorphous solid [1]. Upon decreasing the temperature, the viscosity of the fluid increases dramatically, as well as the time scale for structural relaxation, until the solid forms; concomitantly, the diffusion coefficient vanishes. This process is observed in atomic or molecular systems and is widely used in material processing. Several theories have been developed to rationalize this behaviour, in particular Mode Coupling Theory (MCT), that describes the fluid to glass transition kinetically, as the arrest of the local dynamics of particles. This becomes manifest in (metastable) non-decaying amplitudes in the correlation functions of density fluctuations, which are due to a feed-back mechanism that has been called 'cage-effect' [2].

In colloids, glasses induced by the steric hindrance are also found at large concentration of particles. Interestingly, the hard spheres (HS) system, widely studied theoretically, can be realized using hard neutral colloids. The seminal work by Pusey and van Megen showed that the dynamics of a (polydisperse) HS system arrests [3], inhibiting the crystallization of the system, which is thermodynamically favored at those high densities. This glass transition has been widely analyzed in terms of the predictions from MCT, finding excellent agreement (see Fig 1, and references in [4,5]). The typical two step decay of the intermediate scattering function, or density correlation function, implies a significant decoupling between short time dynamics and structural relaxation. This differentiation of time scales increases as the glass transition is approached, until the structural relaxation (second decay) does not take place (ideally) beyond the critical density. In this case, the system is kinetically arrested, therefore non-ergodic, and behaves like a solid (the elastic modulus grows and overtakes the viscous one). However, in addition to these repulsion driven glasses, equivalent to atomic or molecular glasses, colloidal systems also present low density amorphous solids, known as gels, when attractive interactions are present [6].

Gelation is the formation of a percolating network (typically fractal) of dense and more dilute regions of particles with voids which coarsen up to a certain size and freeze when the gel is formed. It is found in systems with strong short-range attractions, and is a universal phenomenon observed experimentally in many different systems, ranging from colloid-polymer mixtures to protein systems and clays. It is also connected with colloidal flocculation. Because short range attractions (of comparable short range) are not present in atomic systems, gelation is found only in macromolecular systems. In addition to the desirable complete fundamental comprehension of the process, gels have important applications in food industry, stability of commercial colloids or paints. Many detailed experiments have been performed using mixtures of colloids with non-adsorbing polymers, where the attraction strength is set by the polymer concentration, and the range by the polymer size. An effective attraction between the colloids appears due to the depletion of polymers between two particles, when they approach each other beyond the polymer diameter. This effective depletion attraction has been studied theoretically since the 1950's [7].

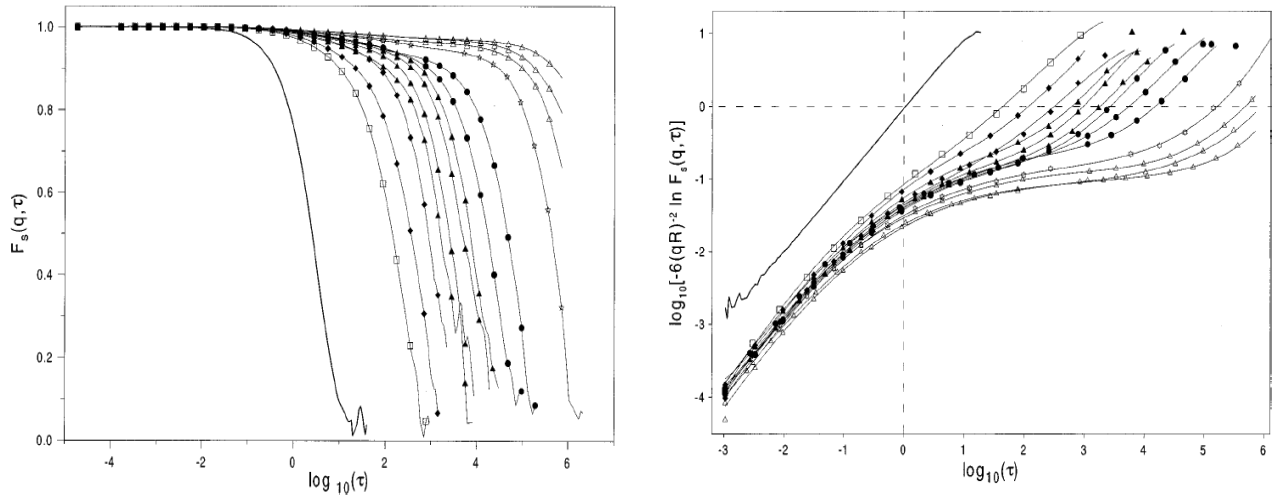

**Figure 1**: *Experimental intermediate scattering function (left) and mean squared displacement (right) for hard colloids, approaching the glass transition. From left to right volume fraction increases from 0.466 to 0.583. The bold line represents the single particle dynamics. With permission from [5].*

In fluids with short range attraction, MCT predicts the existence of two glasses with different driving mechanisms [8,9]. At high density and high temperature, a glass transition induced by core repulsion is found, which is identical to the glass transition found in hard spheres or atomic systems. On the other hand, at low temperatures and all densities vitrification is driven by the formation of long-lived physical bonds between the particles, that hinder their motion. Although this attraction driven glass is exactly in the region where gelation is found experimentally, identification of gelation with it is however not straightforward. At low temperatures, a fluid-fluid transition, equivalent to the liquid-gas transition in atomic systems, is also present – typically this fluid-fluid transition is metastable inside crystal-fluid separation for small attraction ranges. The interplay of the attractive glass transition with fluid-fluid separation poses a major problem for the interpretation of experimental results of gelation.

Figure 2 shows the phase diagram of colloid-polymer mixtures [10,11]. Upon increasing the attraction strength (polymer concentration), the following phases are observed (left pane of Fig. 2): fluid states, phase separation (crystal-fluid or fluid-fluid), non-equilibrium clustering, and finally gelation (non-equilibrium clustering is distinguished from gelation since the former leads to independent non-percolating long lived clusters that sediment). In some cases, however, the transition from the fluid state to the gel occurs directly, without other phase transitions taking place between them (right panel). In all cases, for strong attractions, phase separations are arrested. The interplay between gelation and equilibrium phase transitions is therefore fundamental to explain this phenomenology, but is as yet an open question. Recent reviews are available in the literature, giving the current state of the art where attractive glasses, gels and phase separation meet [12,13,14].

In this chapter, we will present results for gels and attractive glasses mainly from computer simulations, where the effects from phase separation or vitrification can be separated. Specific systems have been devised where the fluid-fluid transition is avoided, or shifted to lower densities, allowing the study of the transition from homogeneous fluids to the attractive glass. At high density, the results will be analyzed using MCT, finding good agreement with the predictions for attraction driven glasses. It will be also shown, however, that the systems are dynamically non-homogeneous, induced by the structural heterogeneity at the particle level, which have been observed by simulations and experimentally. At low density, the decay of the correlation functions at wavevectors around the neighbour peak in the structure factor occurs via joint motions in clusters or flapping of branches, which however, do not cause structural

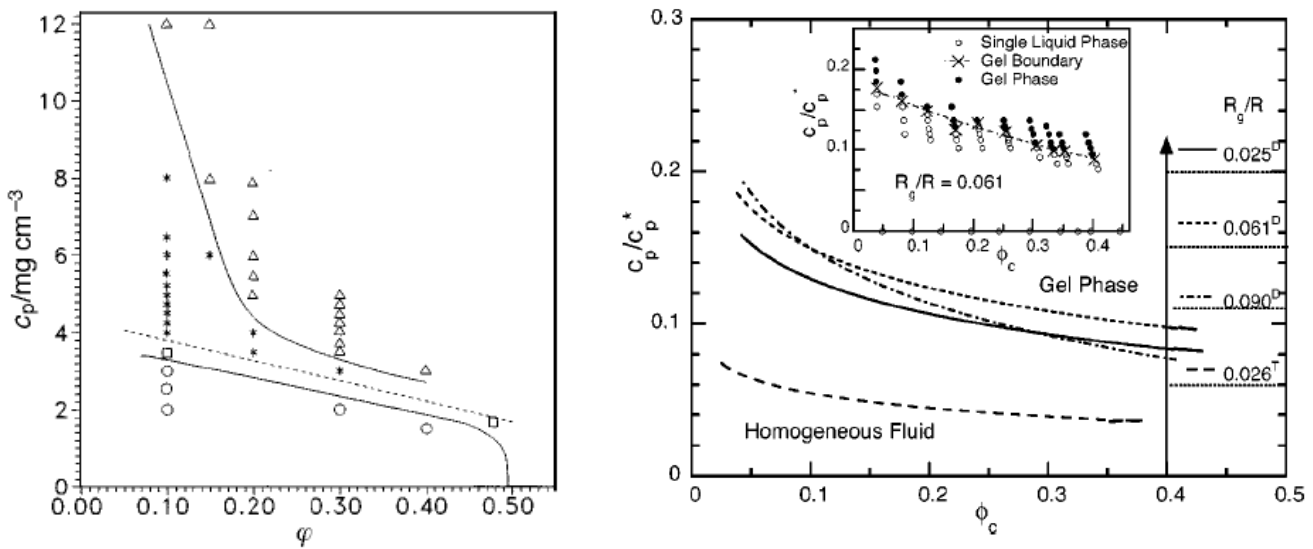

**Figure 2**: *Experimental phase diagrams of colloid-polymer mixtures. The left panel shows fluid states (circles), fluid-crystal coexistence (squares), non-equilibrium non-percolating clustering (asterisks) and gels (triangles) – with permission from [10]. The right panel shows phase diagrams with polymers of different sizes. In the inset, only fluid and gel states are observed – with permission from [11].*

relaxation at long distances.

This chapter is organized as follows: The next section will introduce MCT, giving the basic equations and ideas of the theory. Next, several systems that have been used to study the attractive glass transition and gelation in the literature, will be presented, giving some details. Section 4 will present the main results of this chapter. First the MCT predictions for the attractive glass will be reviewed; the competition of fluid-fluid separation with the attractive glass will be shown next. Then, the MCT analysis of the transition found in simulations in systems without fluid-fluid separation, with the study of the dynamical heterogeneities. Finally, the results at low density and the connection with experimental gels will be discussed. Section 5 will present the main conclusions of this chapter.

## 2. THEORY

In hard-sphere colloidal dispersions, the liquid-glass transition has been studied by dynamic light scattering [15]. Auto-correlation functions for density fluctuations have been measured over about four decades in time. It was found that these correlations decay quickly to zero, as expected for a liquid, only for densities below a critical value. Close to, and especially above, this value, there opens an extended intermediate time window, where the correlators stay (almost) constant at a finite amplitude, see Fig. 1 (right panel). This observation is equivalent to the arrest of the dispersion into an amorphous solid, viz. a glass. The amplitude is the Debye-Waller factor of the amorphous solid, also called glass form factor, and generalizes the order parameter introduced by Edwards and Anderson in the theory of spin glasses. The evolution of the glassy dynamics for the HS system was studied comprehensively by Pusey and van Megen and coworkers [4, 5]. The data suggest that it is the well known cage effect which causes the glassy dynamics and the arrest of density fluctuations in this hard sphere dispersion.

The cage effect is the essential physical concept underlying the MCT for the evolution of glassy dynamics in simple supercooled liquids [16,17]. This theory allows the calculation of the structural dynamics from the equilibrium structure factor [2]. In Refs. [4], detailed quantitative comparisons of the data for hard-sphere colloids with the MCT predictions were presented. It is shown that the theory accounts for the experimental facts within a 15%-accuracy level. The evolution of glassy dynamics was also studied for polymeric micro-network colloids that behave quite hard sphere like [18], and similar conclusions on the validity of MCT were reached. Recently, also the linear and nonlinear rheology of hard sphere like dispersions was compared quantitatively to MCT supporting the conclusions from light scattering [19].

MCT gives a self-consistent equation of motion of the (normalized) intermediate scattering function. This is the autocorrelation function of coherent density fluctuations with wavevector $q$, defined by

$$\Phi_q(t) = \frac{1}{S_q} \left\langle \frac{1}{N} \sum_{i,j}^{N} e^{-i\vec{q}\left(\vec{r}_i(t) - \vec{r}_j(0)\right)} \right\rangle$$

The number of particles is denoted by N. The normalization to unity at time t=0 is provided by the static structure factor

$$S_q = \left\langle \frac{1}{N} \sum_{i,j}^{N} e^{-i\vec{q}\left(\vec{r}_i - \vec{r}_j\right)} \right\rangle$$

which captures equilibrium density correlations. The equation of motion for the correlators takes the form of a relaxation equation with retardation (the retardation or non-Markovian effects are contained in the memory kernel $m_q(t)$ ):.

$$\Gamma_q \, \dot{\Phi}_q(t) + \Phi_q(t) + \int_0^t m_q(t - t') \, \dot{\Phi}_q(t') = 0$$

Without retardation, correlators would show fast diffusive relaxation on the timescale $\Gamma_q = \left( D_s q^2 / S_q \right)^{-1}$ determined by the short time collective diffusion coefficient $D_s$; it captures instantaneous particle interactions, including solvent mediated, so-called hydrodynamic ones, which do not require structural particle rearrangements to act. This decay mechanism will not play an important role at the glass transition.

The central quantity capturing slow structural rearrangements close to glassy arrest is the memory function $m_q(t)$. It can be regarded as a generalized friction kernel, as can easily be verified after Fourier transformation. In MCT-approximation it is given by

$$m_q(t) = \frac{1}{2} \sum_{\vec{k}, \vec{p}} V_{\vec{q}\vec{k}\vec{p}} \Phi_k(t) \Phi_p(t)$$

The vertices $V_{\vec{q}\vec{k}\vec{p}}$ couple density fluctuations of different wavelengths and thereby capture the "cage effect" in dense fluids [2]. It thus enters the theory as a nonlinear feedback mechanism where density fluctuations slow down because of increased friction, and where the friction (precisely the time-integral over the memory kernel which dominates the long time collective friction coefficient) increases because of slow density fluctuations. MCT is a first principles approach as the vertices are calculated from the microscopic interactions

$$V_{\vec{q}\vec{k}\vec{p}} = S_q \, S_k \, S_p \, \frac{\rho^2}{N q^4} \left[ (\vec{q} \cdot \vec{k}) c_k + (\vec{q} \cdot \vec{p}) c_p \right]^2 \delta(\vec{q} - \vec{k} - \vec{p})$$

The mode coupling approximation for $m_q(t)$ yields a set of equations that needs to be solved self-consistently. Hereby the only input to the theory is the static equilibrium structure factor $S_q$ which enters the memory kernel directly and via the direct correlation function $c_q$ that is given by the Ornstein-Zernicke expression $c_q = \left( 1 - 1/S_q \right)/\rho$ with $\rho$ the average density. In MCT, the dynamics of a fluid close to the glass transition is therefore completely determined by equilibrium quantities plus one time scale, here given by the short time diffusion coefficient. The theory thus can make rather strong predictions as the only input, viz. the equilibrium structure factor, can often be calculated from the particle interactions, or even more directly can be taken from the simulations of the system whose dynamics is studied.

The MCT equations show bifurcations due to the nonlinear nature of the equations. A bifurcation point is identified with an idealized liquid-to-glass transition. The quantity of special interest is the glass form factor or Edwards-Anderson non-ergodicity parameter, $f_q$. It describes the frozen-in structure of the glass and obeys

$$\frac{f_q}{1 - f_q} = \frac{1}{2} \sum_{\vec{k}, \vec{p}} V_{\vec{q}\vec{k}\vec{p}} \ f_k f_p$$

In the fluid regime density fluctuations at different times decorrelate, so that the long time limit vanishes. On approaching a critical packing fraction or a critical temperature, MCT finds that strongly coordinated movements are necessary for structural rearrangements to relax to equilibrium. MCT identifies two slow structural processes, β- and α-process, when the glassy structure becomes metastable and takes a long time to relax. In the idealized picture of MCT, the glass transition takes place when the particles are hindered to escape from their neighbouring environments. The non-ergodicity parameter jumps from zero in the liquid to a finite value. This also is accompanied by diverging relaxation times.

Although experiments on molecular glass formers have revealed that the dynamics very close to the transition point is dominated by thermally activated hopping processes, which the described (idealized) MCT cannot account for, MCT has been quite successful in describing the approach to glassy arrest in colloidal dispersions, as mentioned above.

For liquid states close to the glass point and long times the correlator approaches the α-scaling law, where the shape of the scaling or master functions are independent of density or temperature. The α-time scale τ diverges in MCT with an power law, $\tau \sim \varepsilon^{\gamma}$ with

$$\gamma = \frac{1}{2a} + \frac{1}{2b}$$

and thus, τ depends only on the separation ε=(φ-φ$_c$)/φ$_c$ from the critical point. The anomalous exponents $a$ and $b$ follow from the equilibrium structure factor at the transition, and take values around $a \approx 0.3$ and $b \approx 0.6$ for HS. In the vicinity of the critical point, von Schweidler's power-law describes the initial α-relaxation from the non-ergodicity plateau to zero

$$\Phi_q(t) = f_q - h_q \tilde{t}^b \left(1 + k_q \tilde{t}^b\right) + O(\tilde{t}^{3b})$$

where $\tilde{t}$ is a rescaled time, $\tilde{t} = t/\tau$, with $\tau$ following the power law shown above. The coefficients $h_q$ and $k_q$ are called critical amplitude and correction amplitudes, respectively [20]. Von Schweidler's law is the origin of stretching (viz.~ non-exponentiality) in the α-process of MCT. More details about MCT, the asymptotic expansions and the scaling-laws can be found in [2,20,21].

### 3. MODEL AND SIMULATION DETAILS.

In order to model experimental systems with short range attractions different models have been used. Spherically symmetric interaction potentials, such as the simple square well (SW) [22], or the Asakura Oosawa (AO) depletion potential [23], which models the colloid polymer mixture considering that the polymers are ideal [24]. However, due to the short range of these simple potentials, crystallization and fluid-fluid phase separations occur in the same region where gelation is expected, what makes more difficult the interpretation of the data. Therefore, strategies to avoid this equilibrium phase separation have been devised.

Keeping the spherical symmetry of the interaction potential a long range repulsive barrier can be added to the short range attraction, that destabilizes the fluid-fluid separation, and allows the study of fluid states close to gelation even at low densities [25, 26]. Polydispersity is used to avoid crystallization, either a continuous distribution or binary mixtures of particles. Fig. 3 presents an interaction potential based on the AO potential, with a repulsive barrier and a polydisperse system.

Alternatively, non-spherical potentials have also been used, where active spots are located on the particle surface [27], or many body interactions, such as a maximum number of neighbours attracted [28], or combination of both [29]. Using this structural or energetic constraints, the formation of a dense phase is hindered, thus impeding crystallization and liquid-liquid crystallization.

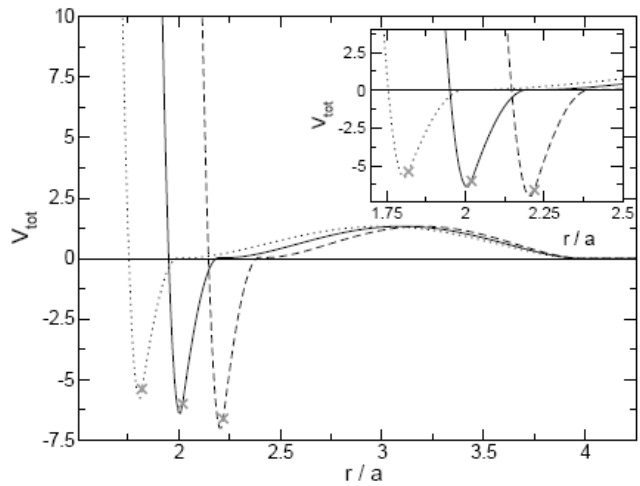

**Figure 3**: *Interaction potential with a short range attraction (given by the Asakura-Oosawa depletion interaction with $\xi = 0.1$) and a long range repulsion of height 1 $k_BT$ (distance is measured in units of the particle radius, a). The different lines correspond to the interaction of particles with different sizes. More details of the interaction potential can be found in [25].With permission from [25].*

Here, we will concentrate on spherically symmetric potentials, which are closer both to the experiments and theoretical developments. For the AO potential , the potential at contact is $V_{min} = -k_B T \phi_p (3/2\xi + 1)$, where $\phi_p$ is the polymer volume fraction and $\xi$ is the ratio of polymer to particle size (which sets the interaction range). The density of colloidal particles is the other key parameter for the phase diagram of the system, which we will report as volume fraction: $\phi = \sum v_i \rho_i$, where the sum runs over populations of particles, $v_i$ is the particle volume and $\rho_i$ is the number density of population $i$ (remember that polydisperse systems are used).

For colloidal systems, the microscopic dynamics for particle $j$ is given by Langevin equation,

$$m\ddot{\vec{r}}_j = \sum_i \vec{F}_{ij} - \gamma \dot{\vec{r}}_j + \vec{f}_j$$

where $\vec{F}_{ij}$ is the interaction force between particles $i$ and $j$, $\gamma$ is the solvent friction coefficient and $\vec{f}_j$ is a random force due to the collisions with the solvent molecules. The friction and viscous forces must obey the fluctuation dissipation theorem: $\langle \vec{f}_j(t) \vec{f}_i(t') \rangle = 6k_B T \gamma \delta_{ij} \delta(t - t')$. It was shown by Gleim et al., that the structural relaxation of a fluid does not depend on the microscopic dynamics [30, 31] as predicted by MCT. Thus, the slow microscopic Brownian dynamics is usually replaced in simulations by the much faster microscopic Newtonian dynamics (in the equation above the friction and Brownian forces are zero), without affecting the structural relaxation.

The dynamical quantities used to study the dynamics of glass transitions are generally the mean squared displacement (MSD), and the intermediate scattering function, or density autocorrelation function, presented above. The MSD is given by

$$\langle \delta r^2(t) \rangle = \left\langle \frac{1}{N} \sum_i (\vec{r}_i(t) - \vec{r}_i(0))^2 \right\rangle$$

where N is the number of particles, and the summation runs over all particles in the system. $\langle ... \rangle$ indicates averaging over time origins. Its Fourier transform is the self part of the density autocorrelation function, $\Phi_q^s(t)$, which is calculated restricting the summation to $i=j$ in the definition of $\Phi_q(t)$ above.

## 4. RESULTS

### 4.1. Theoretical predictions for repulsion and attraction driven glasses.

The theoretical results to be reviewed deal with dispersions of particles whose pair-interaction potential includes a short ranged attraction. The quintessential system, also studied intensely experimentally, is characterized by a hard-core repulsion, and a short ranged attraction potential induced by the depleted polymers. The theory used, namely MCT, was developed in the high-density regime so that the cage effect is essential for the dynamics. Its extrapolation to lower densities is thus quite speculative and requires that solidification is dominated by considerations of local interaction effects like caging and the formation of physical bonds. The relative attraction-shell width is assumed to be small, say less than 10% of the average particle separation. This criterion is based on the Lindemann ratio typically observed at solidification in atomic/molecular systems. There the considerations of packing (viz. the repulsive interactions) dominate solidification and the (possibly present) attraction just affects the adhesion energy and the overall compressibility. Solidification by repulsion commonly allows each particle to explore a region around its site of the size of 10% of the average particle separation. If the attraction range is shorter and if the strength of attraction is sufficiently strong, which is possible in the mentioned colloidal solutions, then crystalline solids with much tighter packing (corresponding to higher density) and amorphous solids dominated by physical bond formation have been observed. Then the particle localization length measured in the mean squared displacement is (far) smaller than predicted by Lindemann, and is of the order of the attraction range. Reviewing the theoretical predictions of this physical bond formation, and its subsequent observation in experimental and simulations studies, is the topic of this chapter.

The main outcome of the theory [8,32,33] is the prediction of two different glassy states, arising because of two different physical mechanisms, on the one hand caging driven by the excluded volume repulsion, and on the other hand physical bond formation driven by the short ranged attraction. The theory predicts a quite complicated diagram of metastable states, glass-to-glass transitions (not yet observed experimentally), and higher-order glass-transition singularities at densities somewhat above the glass transition value of the hard-sphere system. The rich results reflect the interplay of two mechanisms for particle localization, i.e. for the arrest of density fluctuations. Both mechanisms arise because of qualitative features in the only input to the theory, the structure factor S(q), and thus their origin can be traced back (quantitatively) to the underlying particle interactions.

### 4.1.1 The input to MCT: the equilibrium structure factor.

The equilibrium structure factor S(q) is the essential input information required in the MCT equations. While in latter sections we discuss the actual S(q) from simulations (some including a weak long-ranged barrier), in this section we discuss the quintessential effect of a short ranged attraction on S(q) at high density. The interaction potential consists of a hard-core repulsion and an attraction, where we consider both square well (SW) (viz. a constant attraction strength), and the discussed AO depletion attraction. The structure can be specified by three control parameters: the packing fraction $\phi$ of the hard cores, the ratio $-U_a(r=d)/k_BT$ of attraction strength at contact relative to thermal energy, and the relative width δ of the attraction shell compared to the particle diameter d.

The spinodal lines of the particles with an attraction of constant depth (SW system) are shown in Fig. 4 for three representative values of the well width. They specify the divergence points of the compressibility caused by the liquid-vapor transition. The local physics of glass formation according to MCT (caging, bonding) is not affected by the long wavelength fluctuations at the liquid-vapor phase transition, which may only mask the glass lines. MCT can presently address states in the one-component region only, and the complete scenario including phase separation and vitrification is yet unknown.

Figure 5 exhibits structure factors calculated within the mean spherical approximation for states marked by diamonds in Fig. 4 (left panel). The S(q) exhibit a principal refraction peak as known from other simple liquids. It is caused by the hard-core driven excluded volume phenomenon. The high-temperature curves 1 and 2 exhibit peaks, which are only slightly smaller and somewhat broader than the peaks of a HS system at the same densities. The attraction modifies the pair correlations and thus the excluded volume effects, as can be inferred by comparing the curves 1 and 3. Lowering the temperature, the short-ranged attraction causes the particles to move closer, therefore, the peak position shifts to higher wavevector upon cooling. The radial pair distribution function develops a more rapidly varying structure at distances which are multiples of the particle diameter (not shown) [8], and this explains the decrease of the peak height and the increase of the peak wings in *S(q)*.

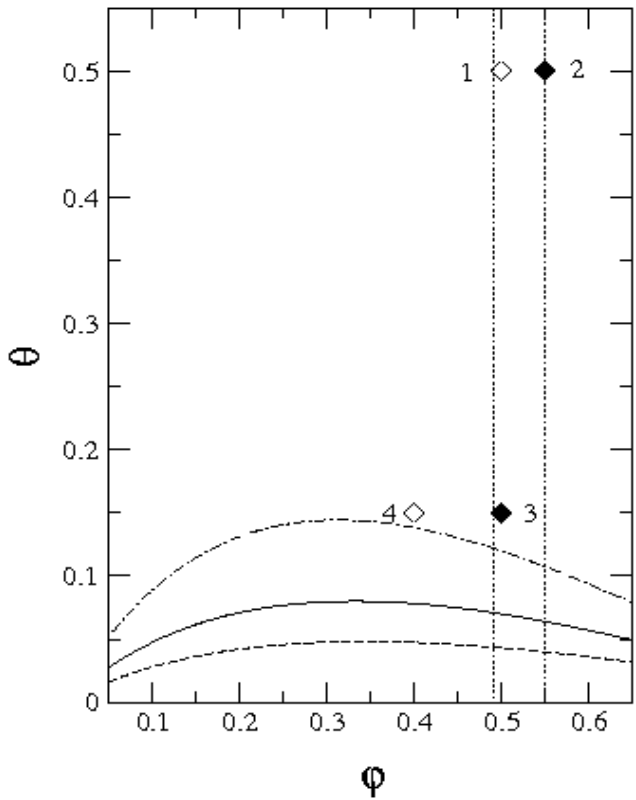

**Figure 4:** *Dimensionless temperature $\vartheta = 1/K = -k_B T / U_a(r/d)$ vs. packing fraction $\varphi$ for the square-well system. The lines show the spinodal calculated within the mean spherical approximation for different relative attraction-well widths: δ=0.03 (dashed line) δ=0.05 (continuous line), and δ=0.09 (dash-dotted). Diamonds mark the state parameters for which structure factors are shown in Fig. 5 (left). With permission from [8].*

The large-wavector tail of *S(q)* will be of importance for physical bonding in the following. In the right panel of Fig. 5 it is shown that a short-ranged attraction causes large and slowly decaying wings extending to very high wavectors; here theoretical and simulations results for the AO potential are shown (colloidal density increases from top to bottom at constant attraction strength) [34].

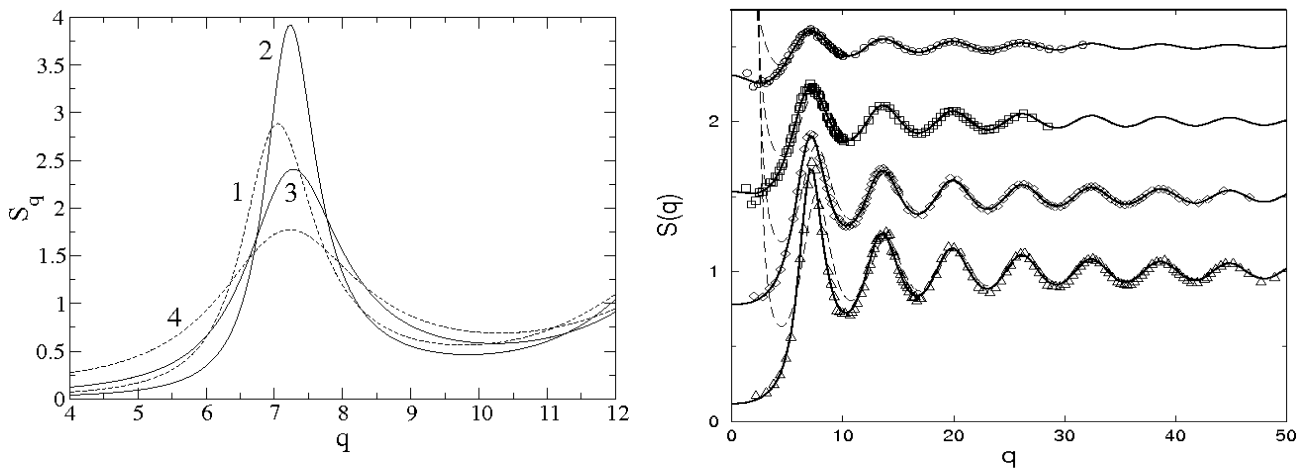

**Figure 5**: *Structure factors as a function of the wavevector q (in units of the diameter) calculated within the mean spherical approximation. Left panel: SW system with relative well width δ=0.05. The labels 1 to 4 correspond to the states indicated by the diamonds in Fig.4. With permission from [8]. Right panel: AO model with $\xi = 0.08$, at fixed polymer concentration of $\phi_p = 0.15$, and varying colloid volume fraction, from top-to-bottom: $\phi = 0.1$, 0.2, 0.3, and 0.4. The symbols are canonical Monte Carlo simulation results and the dashed lines show the simple asymptotic result and the solid lines the full mean spherical approximation. (The curves and data have been offset for clarity; they oscillate about unity in all cases.). With permission from [34].*

MCT finds that the strength of the primary peak in $S(q)$ determines repulsion driven glass formation, while the strength of the large-q tail in $S(q)$ present for short-ranged attractions determines attraction driven glass transitions [35]. Long ranged attractions, which affect the equilibrium phase diagram and the small-q region in $S(q)$ are unimportant for glass formation in MCT [36].

In the presence of a short-ranged attraction, the large-q tail of $S(q)$ generically takes the approximate form

$$S(q) - 1 \approx n\, c(q) \approx \frac{\varphi K}{bq} \left( \frac{\sin q + q/b \cos q}{(q/b)^2 + 1} \right)$$

where $K$ measures the strength of the attraction, and $b \sim 1/\delta$ is inversely proportional to the range of the attraction. This form captures the large-q oscillations, as shown in Fig. 5 (right panel) by the dashed lines. Equivalent expressions have been found for different interaction potentials, using different liquid state approximations, and when working at low and high densities [8,33,34,37]. For an attraction of shorter and shorter range, corresponding to $b \to \infty$, a tail $\sim sin(q)/q$ appears in $S(q)$ which extends to high wavevectors and dominates the glass transition in MCT [35]. It is the origin of bond-formation in MCT.

### 4.1.2 MCT of repulsion and attraction-driven vitrification.

The phase diagram for the SW system is shown in Fig. 6 based on the mean spherical approximation. The two states 1 and 2 from Figs. 4 and 5 are included for δ=0.06. State 1 refers to the liquid phase, increasing $\phi$ to the state 2 increases the height of the first sharp diffraction peak of S(q), and leads to arrest in a glass state, as known from HS. If one cools state 1 at fixed $\phi$=0.50 down to state 3, the primary peak in *S(q)* decreases, yet the large-*q* tail increases. As a result of this compressibility increase on the wings of the structure factor peak, the liquid freezes to a glass upon cooling. For large temperature, *S(q)* depends only weakly on T; this explains, why the transition lines are almost vertical in Fig. 6. On the other hand, the large-q tails in *S(q)* are not very sensitive to density changes. This explains, why the transition lines in Fig. 6 are rather flat as function of temperature.

Repulsion driven glass melts due to cooling, if the decrease of the primary peak in *S(q)* is not overcompensated by the increase of the structure-factor-peak wings. This leads to the slant of the repulsive glass transition line to higher densities upon lowering the temperature, or equivalently increasing the attraction strength. The attraction causes bonding, in the sense that the average separation of two particles is smaller than expected for HS. Therefore the average size of the holes increases and this favors the long-distance motion characteristic for a liquid. This melting of a glass upon cooling occurs only for short ranged attractions, because they distort the local cage. The system remains in the liquid upon further cooling until it reenters the glass when physical bonds become (infinitely) long lived. The described reentry phenomenon (viz. the path glass-fluid-glass possible e.g. at $\phi$=0.525 when changing the temperature in Fig. 6) is a manifestation of the two local mechanisms for localization.

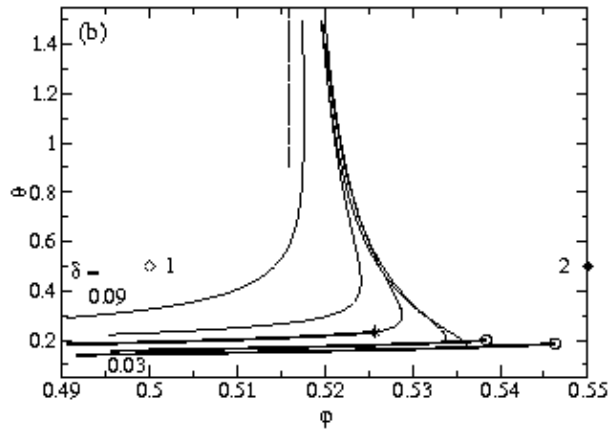

**Figure 6:** *The glass transition lines of the SW system using S(q) from the mean spherical approximation showing cuts through the control parameter space for fixed relative attraction-well width δ; the values are δ=0.09, 0.06, 0.0465, 0.035, and 0.03 (from left to right). Endpoints of higher order MCT singularities are marked by open circles (A₃) and by an asterisk (A₄); for information see [8]. The vertical dashed line marks the transition line for the hard-sphere system, and states 1 and 2 from Fig. 4 are included as diamonds. With permission from [8].*

Figure 7 shows a quantitative calculation for the AO system compared to experimental data from the Edinburgh group [38,39]. The theoretical curves for varying attraction range (proportional to the relative polymer size $\xi = 2R_g / d$), exhibit the melting of the repulsion driven glass upon increasing the strength of the depletion attraction, the almost horizontal line of bond-formation, and the short piece of glass to glass transitions ending in a higher order MCT singularity; it appears only for short enough attraction ranges and is discussed in detail in Refs. [37,40].

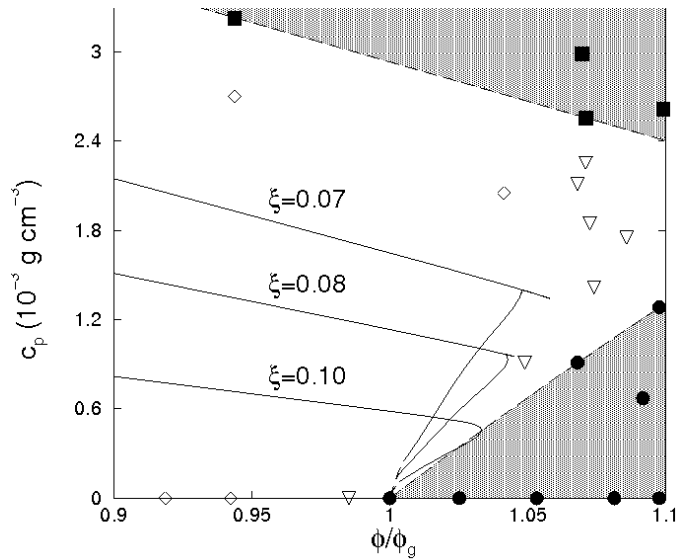

**Figure 7**: *Comparison of MCT predictions (continuous lines) in the colloid-rich part of the phase diagram with experimental data from Pham et al. [38,39]. The lines separate equilibrium states from nonergodic glass and gel states (repulsive and attractive glasses) respectively, in terms of the mass concentration of polymer $c_p$ and the colloid volume fraction $\phi$ relative to the glass transition $\phi_g$ for three values of the polymer-colloid size ratio ξ. The symbols are experimental data: fluid-crystal coexistence (diamond), fully crystalline dispersions (triangle down), repulsion driven glass states (filled circles), and attraction driven glass states (filled squares). The grey regions highlight where non-ergodic states were observed experimentally. With permission from [34].*

The two glass states differ in their local packing and consequently stiffness. While repulsion localizes particles according to Lindemann's criterion, bond formation allows particles less local free volume. Fig. 8 shows the localization lengths (left panel), calculated from the mean squared displacement at infinite time, along the transition lines in Fig. 7.

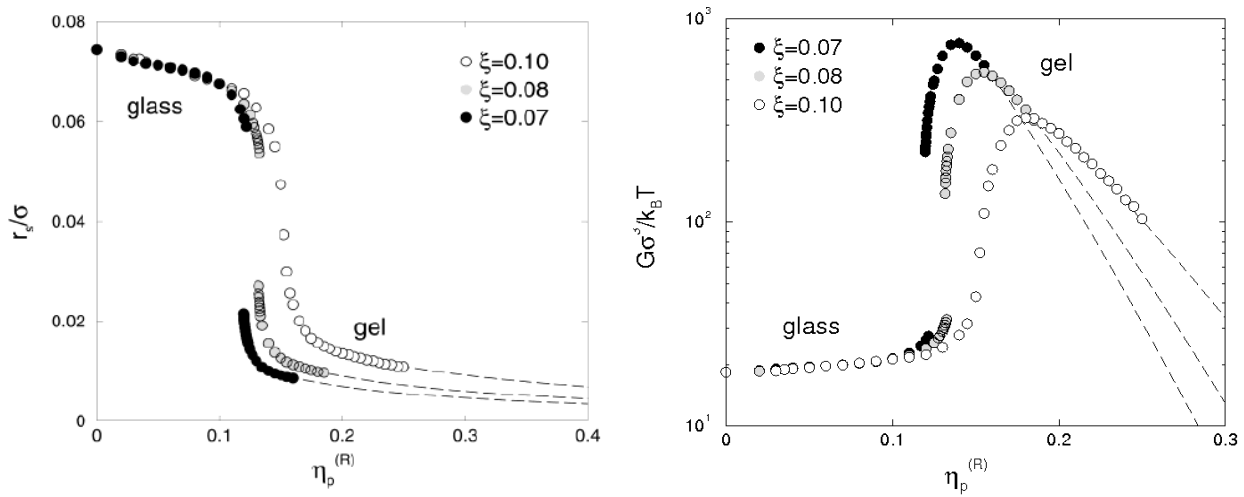

**Figure 8**: *Left panel: Localization length or root-mean-square displacement in the glass/gel state as a function of polymer concentration for three values of ξ along the transition lines in Fig. 7. With polymer concentration the attraction strength increases. The rapid change of $r_s$ is apparent between the Lindemann value characteristic for hard sphere glasses at low attraction strength and the attraction range at large strength where physical bonds are formed. Right panel: The zero-frequency shear modulus as a function of polymer concentration for three values of polymer to colloid size ξ along the glass transitions lines, which are shown in Fig. 7. Note that the particle concentration changes along the curves shown. The dashed lines corresponds to asymptotic results for the small density and attraction range limit [34]. With permission from [34].*

The rapid decrease of the localization length at somewhat higher polymer concentrations, is caused by the effect that the dominant mechanism of structural arrest changes from caging to bonding. For sufficiently small values of $\xi$, the change is discontinuous, and the localization length jumps from higher values in the glass to lower values in the gel. This jump occurs at the crossing of the two glass lines in Fig. 7 (best seen for $\xi = 0.07$). For somewhat larger values of $\xi$ there is a continuous, albeit abrupt, changeover. The typical localization length in the gel state is some fraction of the attraction range, indicating that bonds with lengths set by the attraction are formed among particles.

On crossing a glass transition line the viscous fluid changes to a solid which deforms in an elastic manner. The frequency-dependent storage or elastic shear modulus $G'(\omega)$, which vanishes in the low-frequency limit when the dispersion is in a fluid state, acquires a plateau at low frequencies $G'(\omega \rightarrow 0) = G_\infty$ . The shear modulus $G_\infty$, viz. the elastic constant of the glass, is another quantity that picks up the difference between the glass formation mechanisms.

At low polymer concentration the shear modulus is close to the hard-sphere value. For the shorter-range attractions, the shear modulus jumps to much larger values. The discontinuity lies again at the crossing of both non-ergodicity lines. For somewhat longer-range attractions, there is a steep but continuous increase of the shear modulus as the polymer concentration is increased. The maximum in the modulus arises from the competition between particle packing and bond strength. At low polymer concentrations the high particle concentration imparts elasticity; increasing polymer concentration leads to an increased elasticity because of the increasing bond strength, whereas at sufficiently high polymer loadings in the gel state the particle concentration is so low that the shear modulus eventually begins to decrease.

### 4.1.3 MCT of bond formation.

Formation of physical bonds within the framework of MCT hinges on short-range, local correlations, while the long-range or long-wavelength structure is considerably less important. The content of bond-formation according to MCT can be compared with the familiar Smoluchowski theory of coagulation of charge-stabilized colloids [41]. For the sake of simplicity we restrict the discussion to low concentrations where only isolated pairs of particles need be considered. Two charged colloidal spheres interact via the Derjaguin-Landau-Verwey-Overbeek potential, which exhibits a deep and narrow primary minimum near contact as well as a repulsive barrier at somewhat larger separations, caused by the interplay between dispersion and screened Coulomb interactions. The aggregation is described as an activation problem, i.e. whether the particle pair can overcome the repulsive barrier and enter the primary minimum near contact. The attraction strength here is large compared to thermal energy. In the Smoluchowski theory of aggregation (doublet formation in the present context) a non-equilibrium situation is postulated at the outset by the choice of the value of the radial distribution function at contact, which differs from the value obtained from the interaction potential via equilibrium statistical mechanical averaging. While equilibrium theory predicts a finite, actually a rather high, contact value, for aggregation its vanishing is enforced because two particles in contact are interpreted as a doublet, and thus drop out of the description provided by $g(r)$. A non-equilibrium flux of particles forming doublets upon overcoming the barrier and reaching contact is established, while kinetic stabilization is regulated by the Coulombic barrier height.

This familiar Smoluchowski picture should be contrasted with the situation addressed by MCT: bond-formation for dilute dispersions of particles interacting via the depletion interaction (or another short-range attraction) of only moderate strength. Since the attraction strength is of order a few $k_BT$ the loss of equilibrium cannot be assumed a priori, rather the theory itself must deliver a criterion for when the attraction strength suffices for the formation of long-lived particle aggregates. Thus the challenge is to find within an equilibrium statistical mechanics approach the threshold interaction beyond which the system falls out of equilibrium, and where non-equilibrium type of approaches, like diffusion limited cluster aggregation, start to become meaningful.

The discussion of the large-q tail in *S(q)* in Sect. 4.1.1, which is characteristic for a short-ranged attraction, enables one to formulate a simplified theory of bond formation within MCT with the result that the long-time limit of the dynamic structure factor is controlled by a single interaction parameter, $\Gamma = K^2 \phi / b$. Bond formation occurs at $\Gamma_c = 3.02\ldots$ [34]. For small values of $\Gamma$, the dynamic structure factor decays to 0 for all wavevectors. Physically, this means that concentration fluctuations decay into equilibrium at long times, just as expected for a colloidal fluid. However, for $\Gamma \geq \Gamma_c$ the solutions yield a non-zero glass form factor, viz. the system arrests in a metastable state. This simple result requires the approximate expression for *S(q)* given above, and needs to be replaced by a full numerical solution whenever this approximation fails.

Figure 9 shows that sufficient polymer concentration, viz. attraction strength, suppresses crystallization at a non-equilibrium boundary, beyond which different types of aggregation behavior are observed. Crystallization and fluid crystal coexistence would be the behavior expected from the equilibrium phase diagram, with fluid-fluid separation metastable, as shown by the critical point (open squared) calculated with the PRISM theory [42]. Immediately across the non-equilibrium aggregation boundary, disordered clusters of colloids are formed in the Edinburgh system that rapidly settle under the influence of gravity to give amorphous sediments. Deeper inside this non-equilibrium region, a rigid gel forms, which settles suddenly after an induction time. Samples in this "transient gel" region not only exhibit a transient rigidity, but also nonergodic dynamics. Yet, MCT is in nearly quantitative accord with the non-equilibrium aggregation boundary, along which crystallization stops, instead of the transient gelation boundary, which is the experimental non-ergodicity line. The theoretical prediction is sensitive to the value of attraction range but even on treating ξ as a freely adjustable parameter, the theoretical line cannot be brought in agreement with the transient gelation boundary. This observation is one of the indications that at low concentrations the role of the MCT bond formation transition with respect to colloidal gelation is yet unclear. The mesoscopic behavior, formation of clusters and/or ramified solids appears to require additional theoretical considerations. Non-ergodicity of the local dynamics as described by MCT, however, appears still a necessary ingredient for non-equilibrium phenomena, as may be concluded from the accord with the non-equilibrium aggregation line in experiment. More information can be gained from simulations as will be discussed in the reminder of this review.

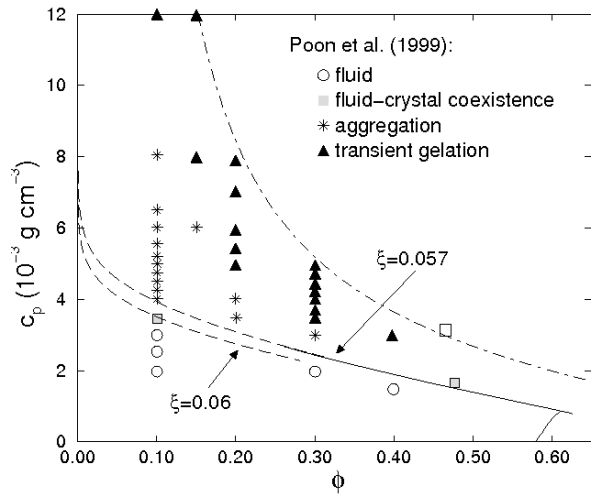

**Figure 9**: *Comparison of MCT predictions with the experimental phase diagram of Ref. [10] (see Fig. 2) in terms of the mass concentration of polymer $c_p$ and the colloid volume fraction for ξ≈0.06. The analytical MCT predictions are shown as broken lines, while a numerical MCT solution is shown as solid line. Also shown is the PRISM prediction for the metastable fluid-fluid spinodal with an open square marking the location of the critical point [42]. With permission from [34].*

## 4.2. Simulations of attractive glasses and gels

Attraction driven glasses as predicted by MCT should extend into the same region as fluid-fluid (liquid-gas) separation, and therefore a competition between both processes can be found. In fact, the current understanding of most experimental gels is based on the arrest of the spinodal decomposition, which complicates the complete rationalization of the vitrification process induced by attractive interactions. In the following, we will show how this arrest of the demixing takes place, and several models where the fluid-fluid separation is inhibited thermodynamically will serve as benchmarks for the theoretical predictions. These models will show that the glass transition is found in the region predicted by MCT, with the correct properties at high density, and that the high order singularity indeed exists. The extension of the glass transition to low density, however, presents new features that are currently absent in the theory. Two snapshots of gels at high and low density are presented in Fig. 10.

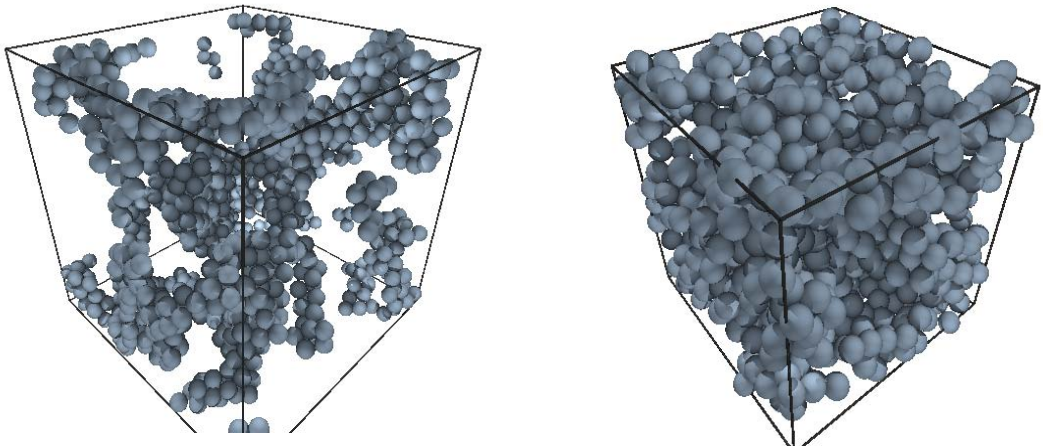

**Figure 10:** *Snapshots of the AO system with arrested phase separation, at volume fraction $\phi = 0.10$ (left) and $\phi = 0.40$ (right), with attraction strength equal to $16k_BT$.*

### 4.2.1. Competition of liquid-gas separation with attraction driven vitrification.

Fluid-fluid phase separation can be monitored using the structure factor at low wave vectors, or by some specific parameter which measures the macroscopic density fluctuations at long times, such as the standard deviation of the density in subsystems. Figure 11 shows the evolution of the structure of the AO system ($\xi = 0.10$) for different attraction strengths (the critical point in this system is at $\phi_p \approx 0.29$), with the time elapsed since the quench, waiting time $t_w$. It is expected that upon increasing the attraction, the fluid-fluid separation boundary is reached and the system separates in two phases with more and more different densities.

For large attractions, however, the separation is prevented, as noticed by the arrest of the low wavevector peak in the structure factor, or by the decrease of the inhomogeneity parameter. The structure of the system is nevertheless not homogeneous, but resembles a network of particles with branches and voids, as can be observed in the snapshots presented in Figure 10 for two

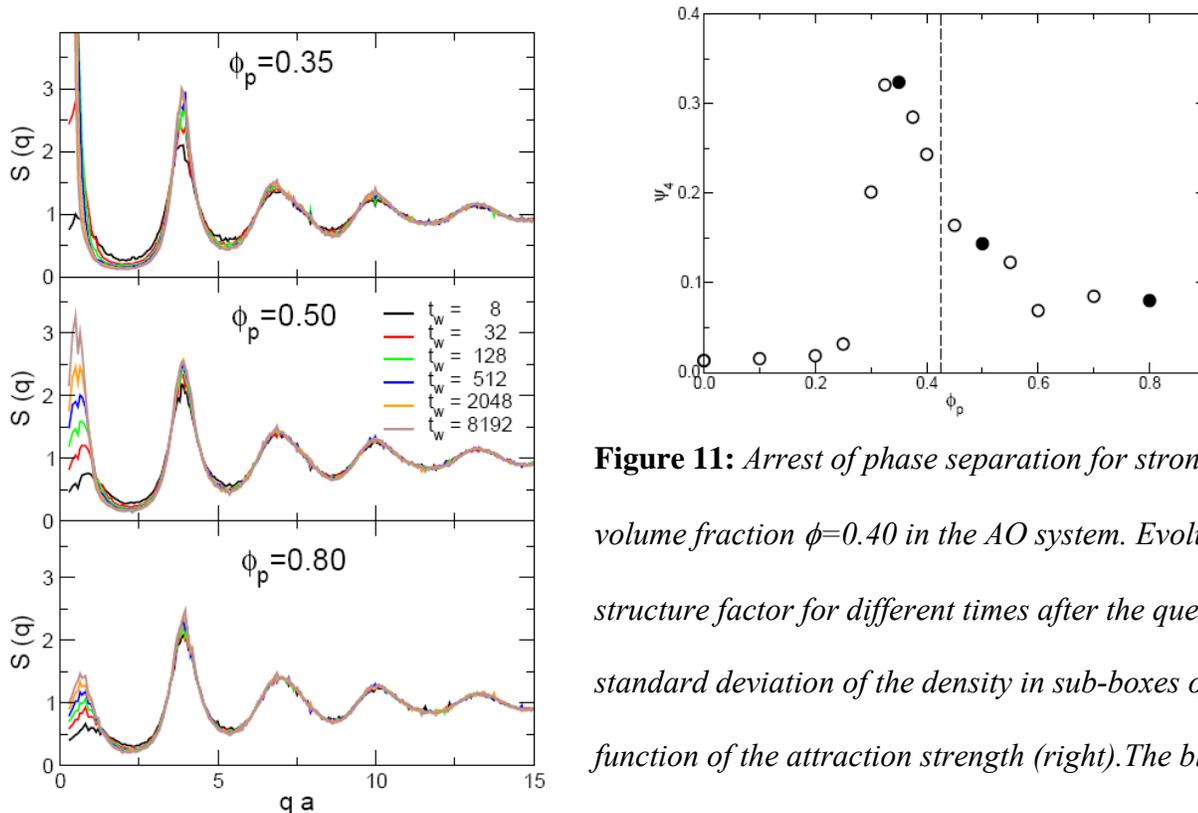

**Figure 11:** *Arrest of phase separation for strong attractions at volume fraction φ=0.40 in the AO system. Evolution of the structure factor for different times after the quench (left), and standard deviation of the density in sub-boxes of size L/4 as a function of the attraction strength (right).The black circles mark three states in the left panel. With permission from [23].*

different volume fractions [23].

The arrest of the phase separation implies obviously an arrest of the dynamics of the system, as can be observed using the self part of the intermediate correlation function, $\Phi_q^s(t)$. In Figure 12, $\Phi_q^s(t)$ is presented for the AO system for different waiting times after the quench, for the three states at $\phi = 0.40$ presented in Figure 11. The system which undergoes equilibrium phase separation at $\phi_p = 0.35$ reaches a stage where the dynamics is independent of the waiting time, and the structural relaxation occurs via diffusion of dense fluid clusters and/or by the exchange of particles between the liquid and gas phases. On the other hand, the systems quenched at states with stronger attractions slow down continuously, without saturating. The evolution of $\Phi_q^s(t)$ with waiting time does not change the height of the plateau, since the structural evolution from the homogeneous fluid to these high density gels only requires local rearrangements (see the structure factor), but only increases the relaxation time (following a power-law with the waiting time, as shown in the lower panel). The correlation functions can be time-rescaled to collapse onto a master decay, what allows an unambiguous determination of the relaxation time [23, 43]. For stronger attractions, the localization length decreases, as deduced from the increase of the non-ergodicity parameter (upper panel in the right), in agreement with states deeper in the glass phase. These results indicate that the dynamics of the system is similar to glass aging.

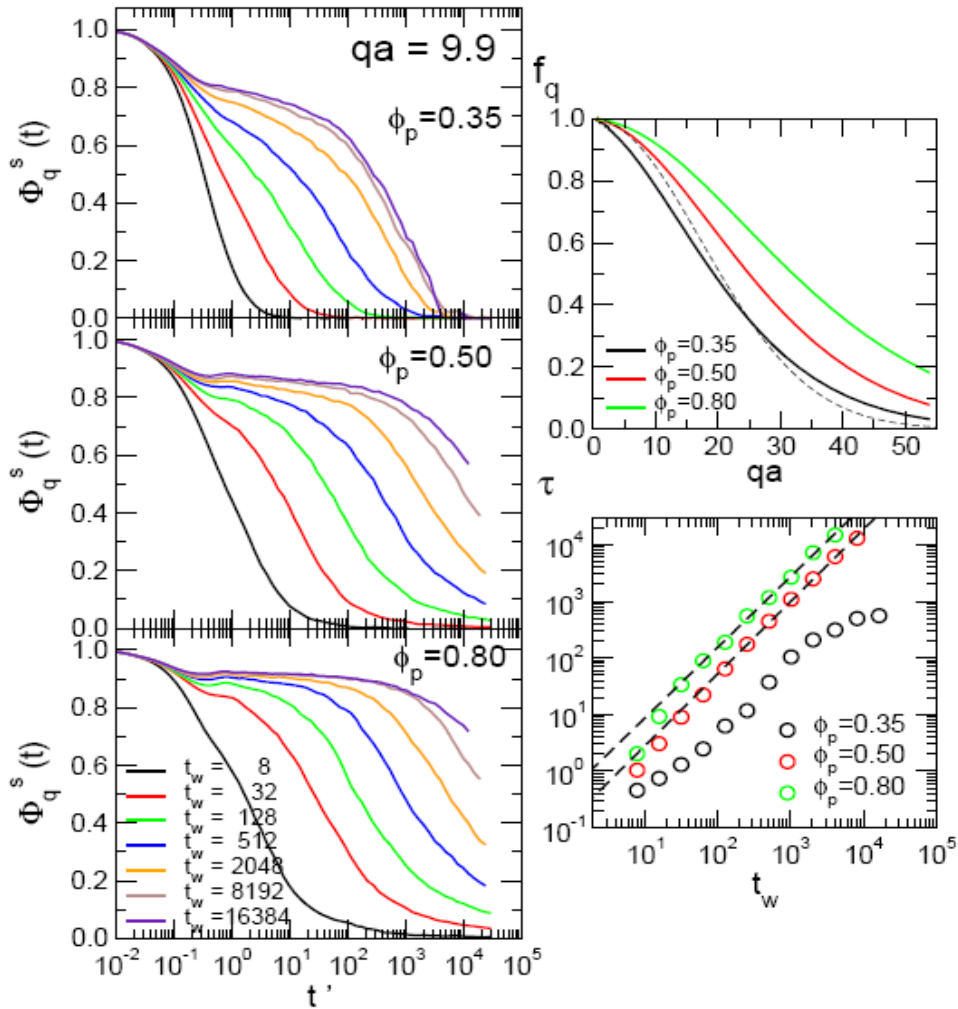

**Figure 12**: *Self part of the density correlation functions for the three circles in Fig. 11, for different waiting times as labeled (left).The panels in the right show the non-ergodicity parameter (upper panel), where the dashed line is a Gaussian fit giving the localization length for $\phi_p = 0.50$, and the relaxation time (lower one) for the three states, with power law fittings for the two states with stronger attractions. With permission from [23].*

At lower density, on the other hand, the structural evolution of the system is more dramatic.

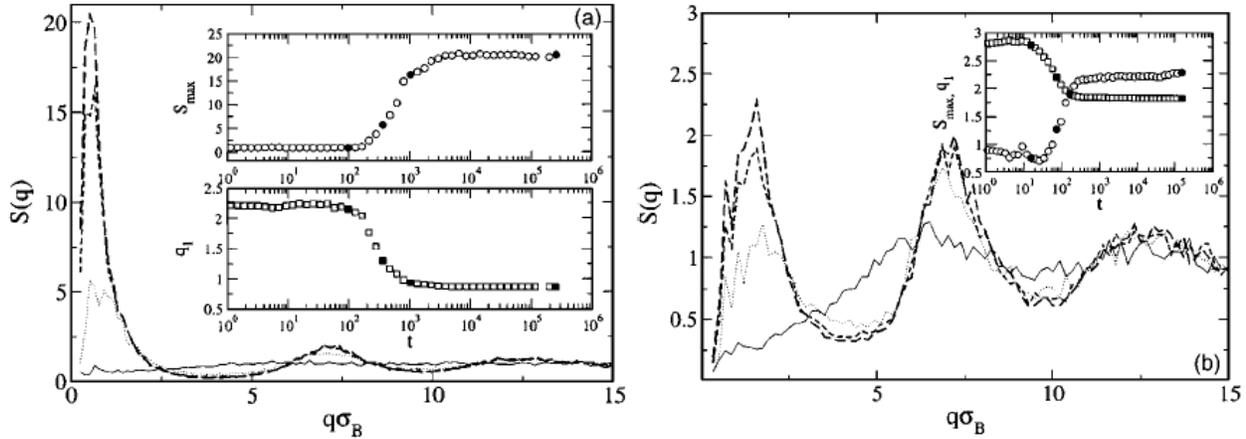

**Figure 13**: *Evolution of the structure factor for a SW binary mixture (50/50 with size ration 1.2) quenched at temperature T=0.05 (the energy scale is set by the well depth) for volume fraction $\phi = 0.10$ (left panel) and $\phi = 0.25$ (right panel). The insets show the evolution of the first maximum height $S_{\max}$ and its position $q_1$ with waiting time, and the closed symbols indicate the times where the S(q) in the main panel is shown. With permission from [44].*

In Fig. 13 the evolution of the structure factor is presented for a SW mixture at low temperatures [44]. The change in S(q), until it finally stops evolving, is more pronounced for lower density. This implies that the structure is much more heterogeneous at lower density than at higher ones, as shown in the snapshots in Fig. 10. In other words, the fluid-fluid separation has been arrested at a later stage, and affects more strongly the dynamics of the system.

In order to avoid the difficulties associated with the structural heterogeneities at low density, the attractive glass transition has been investigated at high density in systems or states where the fluid-fluid transition is not present. In section 4.2.5. we will go back to the problem at low density.

### 4.2.2. Simulations of attractive glasses.

In order to avoid the possible effects from the fluid-fluid demixing, different strategies have been used to reach the attractive glass transition from fluid states (crystallization is always avoided using a binary or polydisperse mixture of particles). One possibility is to study the attractive glass in systems with very high density, larger than the denser fluid in the phase separation. On the other hand, interactions potential that hinder phase separation, keeping the short range attraction, have been used.

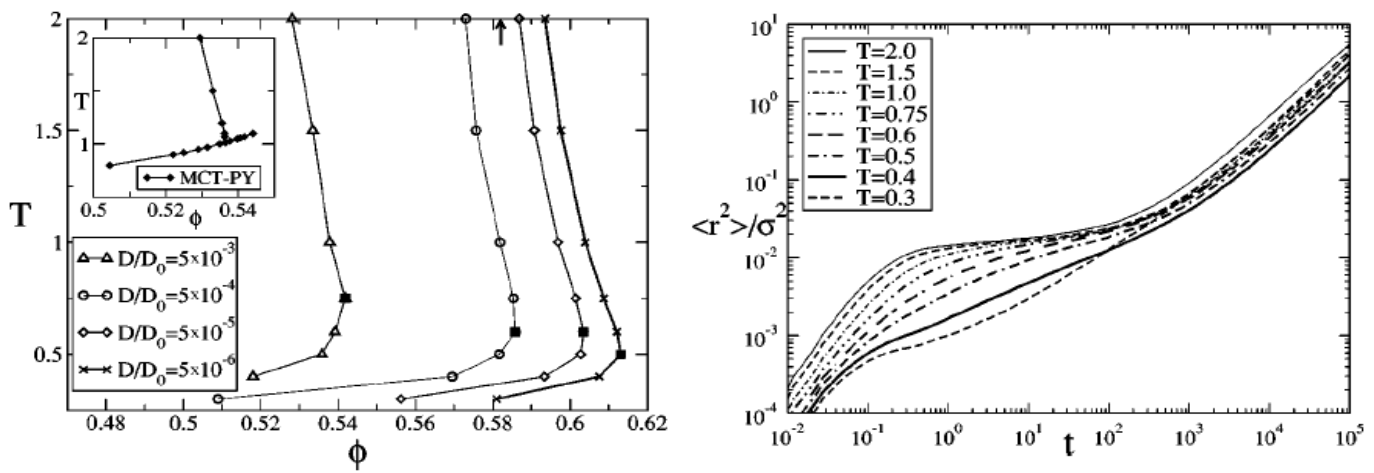

**Figure 14**: *Attractive glass in the SW system. Iso-diffusivity lines (left) finishing in the liquid-gas spinodal (the inset shows the MCT prediction for the same system), and the MSD for different states along the iso-diffusivity line with lowest D (right). With permission from [22].*

The left panel of Fig. 14 shows the iso-diffusivity lines -lines joining states where the diffusion coefficient is constant- for the SW mixture of Fig. 13 in the high density region up to the liquid branch of the spinodal [22]. The overall shape of the lines, and its extrapolation to zero diffusion coefficient, follows the theoretical predictions by MCT (inset), particularly in finding the fluid pocket between the attraction driven (low T) and repulsion driven (high T) glasses. The mean squared displacement along an iso-diffusivity line shows that the trapping distance is different at low or high temperatures, signaling the different origin of both glasses (steric hindrance for the repulsive glass and bonding for the attractive one).

These results, however, are restricted to high densities, beyond the liquid branch of the fluid-fluid separation. In order to study the transition at lower density, it is necessary to prevent the phase separation. Figure 15 presents the MSD for different states along the isochore $\phi = 0.50$ for the AO system with the repulsive barrier presented in Fig. 3. Similar to the previous figure, this one shows the reentrant glass (note the disappearance and reappearance of dynamical slowing down as the attraction strengthens), with different driving mechanisms for weak and strong attractions [38]. The density is lower in this case than in Fig. 14, since the repulsive barrier inhibits the fluid-fluid separation. This shows that the rich scenario predicted by MCT does not change qualitatively due to the long range repulsion added to the system, and is unaffected by the details of the attraction.

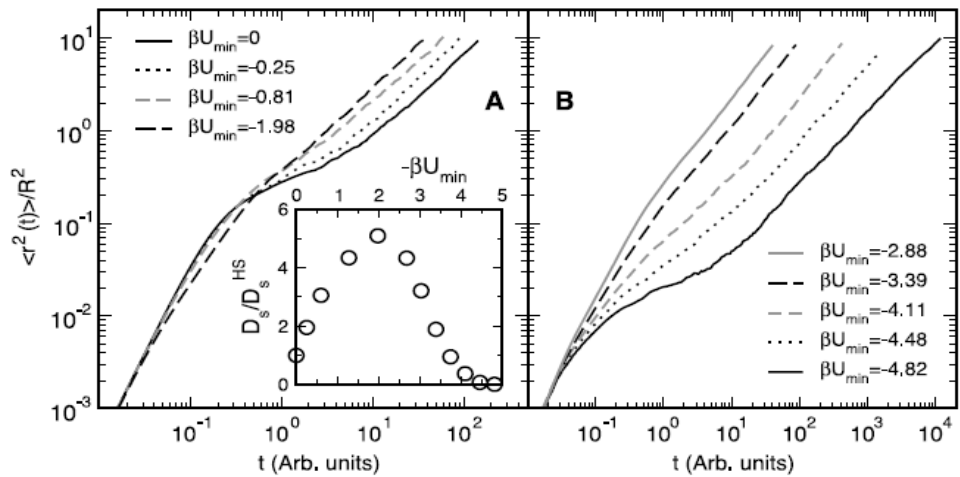

**Figure 15:** *MSD for the AO system with repulsive barrier, for different states with volume fraction $\phi = 0.50$ and the attractions strength as labeled. The inset shows the diffusion coefficient from the long time slope of the MSD. With permission from [38].*

This system, therefore, can be used to analyze the properties of the glass transition reached from the fluid side, without effects from the phase transition at moderate density. Before studying the dynamics of the system, it is necessary to study its structure to answer the question: How does the system avoid the phase separation? The structure factor is presented in Fig. 16 for different states along the isochore $\phi = 0.40$. A clear (pre-) peak appears at low wavevectors, indicating the formation of holes in the system, and denser regions elsewhere. This peak shows the tendency of the system to undergo microphase separation [45, 46], although for the states in the Figure the peak is too small to claim that this separation has occurred. Additionally, the increasing amplitude of the oscillations at large wavevectors signals the local reordering due to the short ranged attraction (compare with Fig. 5) [25].

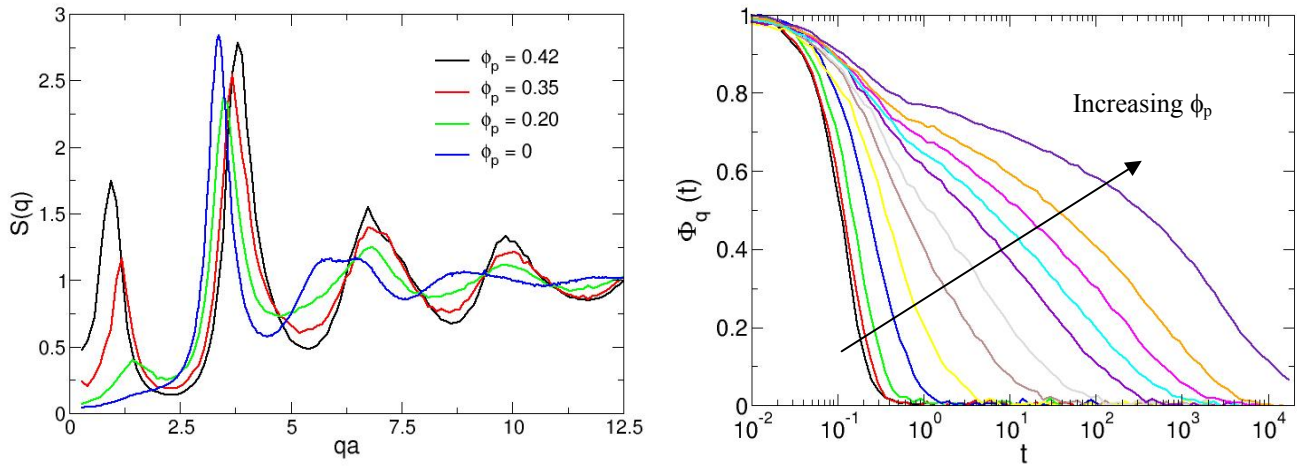

**Figure 16:** *Structure factor of the AO system with repulsive barrier for different states with $\phi = 0.40$, as labeled. The panel in the right presents the density correlation functions for states in the same isochore for the third peak in $S(q) - qa=9.9$ (a is the particle radius) from $\phi_p = 0$ up to $\phi_p = 0.42$.*

The density correlation functions of this system for different states increasing the attraction up to the glass transition is presented in the right panel of Fig. 16, for the wavevector *qa=9.9*, which corresponds to the third peak in *S(q)*. A two-step decay, typical of glass forming systems is observed for high attractions, with an important stretching of the second decay. The height of the plateau, the non-ergodicity parameter, is much larger than for the repulsion driven glass (comparison shown in Fig. 18), in agreement with the shorter localization lengths of attractive glasses with respect to repulsive ones (see Fig. 8).

The analysis of fluid states close to the glass transition can be performed using the predictions from MCT from Sect. 2. Fig. 17 presents the von Schweidler fittings to the correlation functions (left panel), and the increase of the time scale for structural relaxation, viscosity, and diffusion coefficient, fitted according to power law divergences (right panel). The von Schweidler exponent, which measures the stretching of the structural decay is lower than for HS, as predicted by MCT: $b = 0.37$ as compared to $b = 0.53$ for HS [25,47]. With this value of *b,* the exponent for the power law divergence is, within MCT, $\gamma = 3.44$, which agrees with the value found from free fitting for the time scale, $\gamma = 3.23$, but disagrees with the value for the diffusion coefficient, $\gamma = 1.33$; a disagreement also found in other (repulsive) glasses [48].

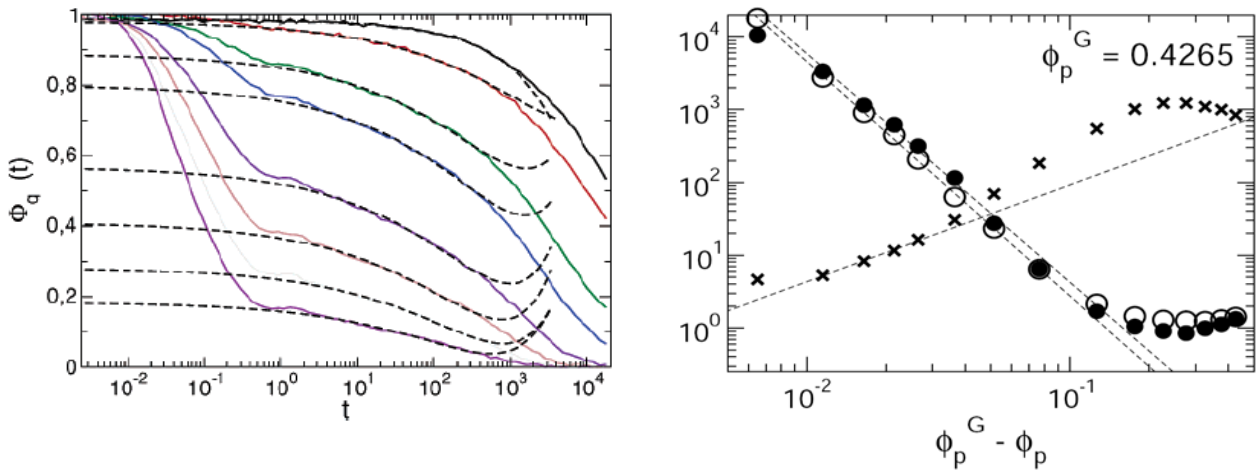

**Figure 17:** *MCT analysis of fluid states close to the attractive glass. The panel in the left shows the Von Schweidler fits at different wavevectors (q increases from top to bottom) for the state* $\phi = 0.40$ *and* $\phi_p = 0.42$. *With permission from[47]. The panel in the right shows the structural relaxation time scale (white circles), the viscosity (black circles) and the self diffusion coefficient (crosses) as a function of the distance to the transition. With permission from [50].*

The increase of the time scale for structural relaxation is also noticeable in the stress correlation function, i.e. the time scale for relaxation of local stresses also increases. As stated by the Green-Kubo relation, the viscosity can be calculated as the time integral of this stress correlation function [49], and is shown in Fig. 17 (right panel) as the transition is approached. A power law divergence is found, with the same exponent as for the density correlator time scale, $\gamma = 3.14$, again in agreement with the MCT prediction [50, 51]. The Fourier transform of the stress correlation (multiplied by the frequency and imaginary unit) gives the elastic and shear moduli. The elastic modulus shows a plateau appearing at low frequencies, and the viscous one develops a maximum followed by a minimum also at low frequencies [50,51], in agreement with the predictions from MCT but also with experimental results of attractive glasses [52]; they can be compared with corresponding measurements in the HS system [19].

A more quantitative comparison can be performed taking the structure factor from the simulations and using it as input of MCT -- the system has to be made monodisperse to provide the correct input for the conditions of the theory, but the repulsive barrier is present. In Fig. 18 the non-ergodicity parameter and time scale for structural relaxation calculated with MCT are compared with direct measurements in the simulations, and the effects of polydispersity and the repulsive barrier tested: (A) corresponds to results for a monodisperse system without barrier,

(B) to a monodisperse one with barrier, and (C) to the polydisperse one with barrier. Semi-quantitative agreement is found in these properties with no adjustable parameters when a monodisperse system is considered, showing that MCT correctly captures the properties of the transition and that the long range repulsion does not affect it. For comparison, the non-ergodicity parameter from simulations of HS is also included (black points), showing the big difference between $f_q$ in the attractive and repulsive glasses. However, the transition point is overestimated by a factor close to two ( $\phi_p^c = 0.426$ in the simulations and $\phi_p^c = 0.246$ in MCT), following the typical trend of MCT of overestimating the driving force for the transition [48]. The von Schweidler exponent is also incorrectly calculated in the theory, probably due to different distance from the transition point to the high order singularity (where the exponent $\gamma$ diverges, and thus $b$ goes to zero) [53]. The time scale for structural relaxation (right panel) shows that the slowest modes to relax are those connected with the pre-peak in S(q), although the transition is not driven by them. Due to the wrong determination of the exponents within the MCT

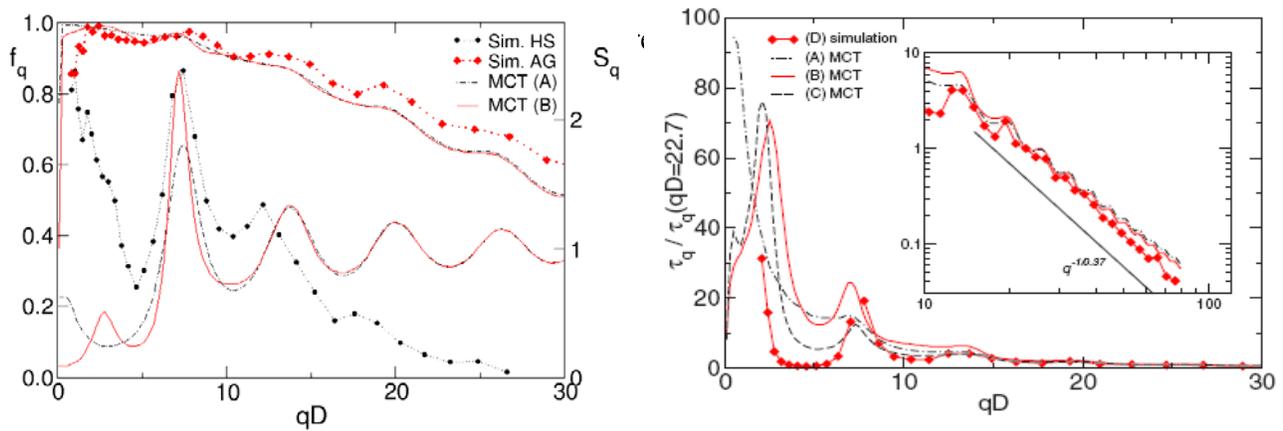

**Figure 18:** *MCT analysis of fluid states close to the attractive glass. With the structure factor from simulations of the system, the non-ergodicity parameter (left panel) and time scale (right panel) from the von Schweidler analysis are compared with the theoretical calculations (continuous lines). The structure factors are included in the left panel, to be read in the right scale. (A) refers to monodisperse system without the repulsive barrier, (B) with the barrier and (C) polydisperse. The non-ergodicity parameter for HS is also included for comparison. With permission from [53].*

Alternatively, the fluid-fluid separation has been avoided using models with a maximum number of interacting neighbours, $n_{max}$ [54]. Fig. 19 shows the phase diagram of this system for $n_{max}=4$, with several iso-diffusivity lines. Note that the liquid-gas separation is restricted to low density, and the liquid branch of the spinodal has a volume fraction below $\phi = 0.30$ for this value of $n_{max}$, which allows the fluid phase to extend to low temperatures. In this system, however, the diffusion coefficient decreases with temperature, according to the Arrhenius law (right panel), showing that the glass transition driven by the attractive interactions is found here only at zero temperature [54]. This proves that a large number of bonds per particle must be formed in order to completely arrest the local dynamics of the particle at a finite temperature; this is achieved in systems where a larger number of neighbors is allowed, or where a dense liquid phase forms.

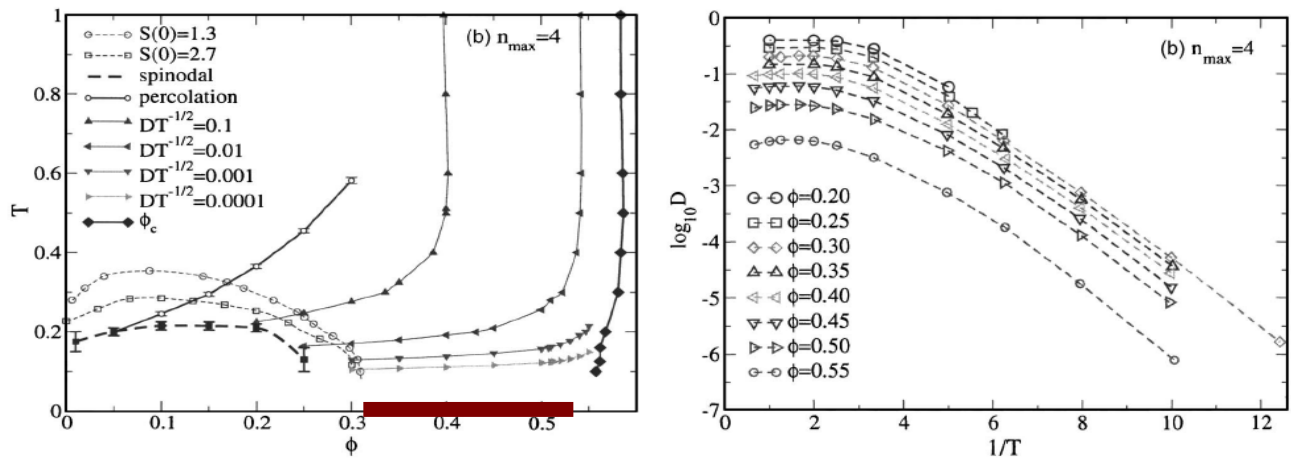

**Figure 19:** *Analysis of the limited valance model with $n_{max}=4$. Isodiffusivity lines with the spinodal (left panel), and dependence of the diffusion coefficient along the isochores indicated as a function of the inverse temperature (right panel). The thick line at the bottom of the left panel shows the expected glass line from the Arrhenius plots. With permission from [54].*

### 4.2.3. Dynamical heterogeneities.

Although the results presented in the previous section show that MCT provides a good description of the attraction driven glass, and even anticipated its main properties, MCT considers the system dynamically homogeneous and does not describe heterogeneities at the particle level. Dynamical heterogeneities of increasing importance as the transition point is reached have been described in simulations of repulsion driven glasses as particles of increased mobility that cluster forming movable regions [55].

In attraction driven glasses, dynamical heterogeneities are also present, as can be seen in Fig. 20, where the distribution of squared displacements is shown for different times for four states close to the glass transition in the AO system with repulsive barrier. Note that upon increasing the attraction strength, the system becomes very heterogeneous from the dynamical point of view; the structural relaxation occurs in widely different time scales for different particles, and two types of particles can be recognized for the closest state to the transition: *fast* and *slow* ones, according to the two maxima in the distribution of displacements in the figure [56,57]. This binary distribution of relaxation times has also been observed in other gel-forming systems [58, 59].

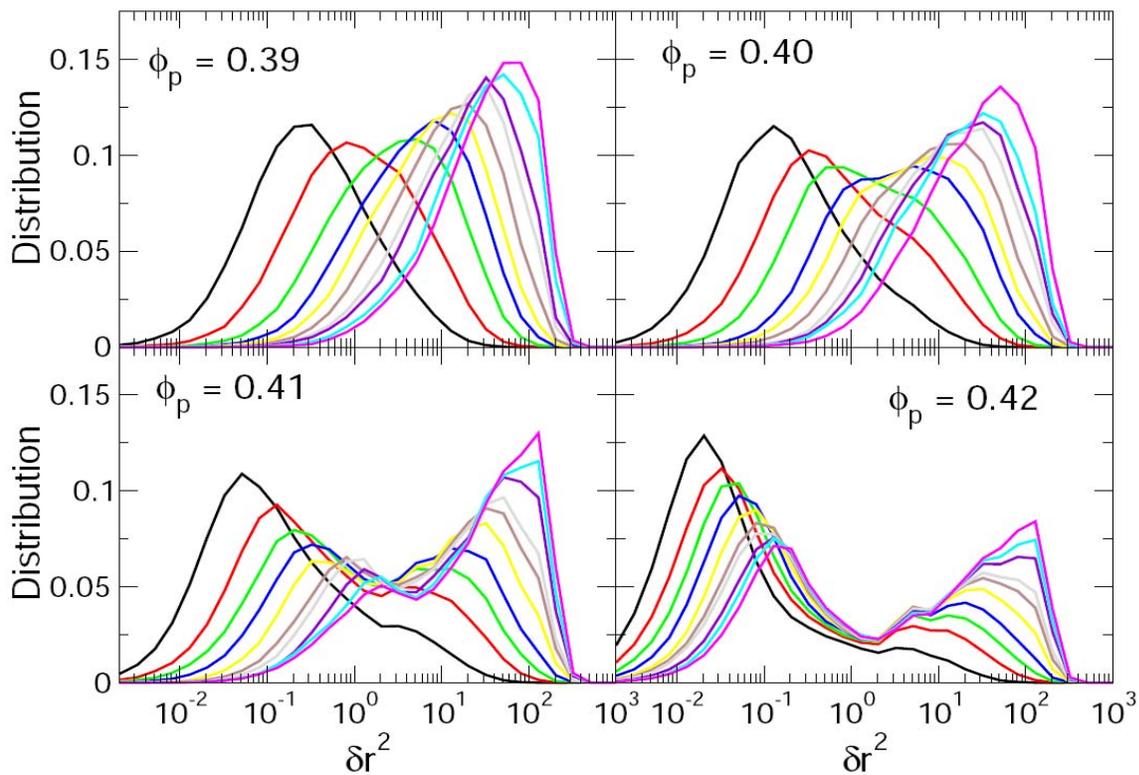

**Figure 20:** *Analysis of the squared displacements of a system close to the attractive glass in the AO system with barrier. The distributions for different states with volume fraction $\phi = 0.40$ as labeled, for different times, increasing from left to right (the attractive glass for this system is at $\phi_p = 0.4265$). With permission from [47].*

Detailed analysis of the systems shows that the particles that take longer to diffuse are in the inner parts of the gel, i.e. are forming the *skeleton* of the network, whereas the fast ones are in the outer parts, forming the *skin*. The structural heterogeneity of the system is therefore responsible of the dynamical one, in contrast to repulsive glasses, where the structure is homogeneous.

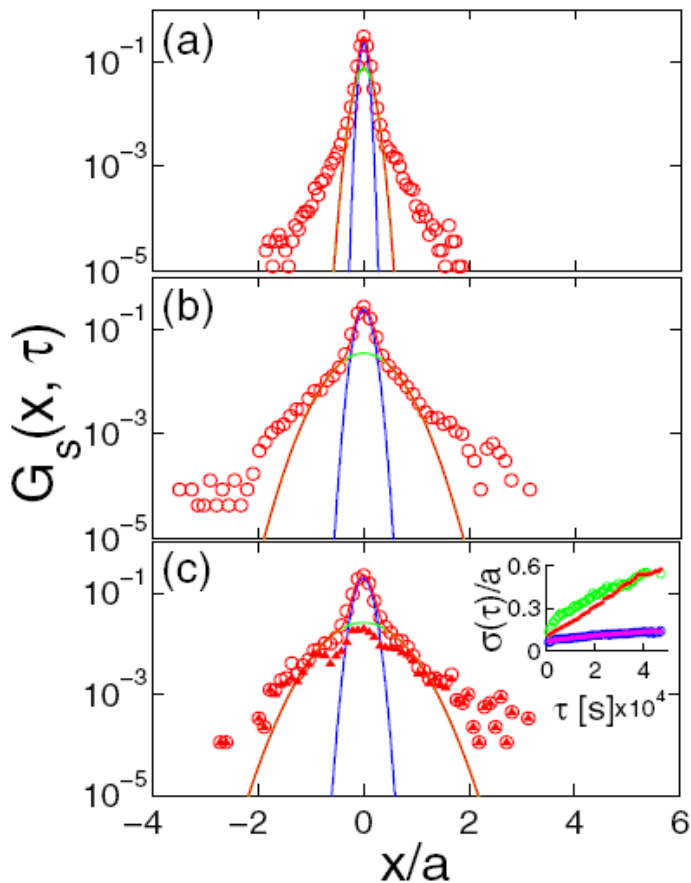

**Figure 21:** *One dimensional distribution of particle displacements at different times increasing from (a) to (c). Slow and fast populations can be recognized from the Gaussian fits to the data (blue for the slow and red for the fast ones). The inset in the lower panel shows the width of both Gaussians as a function of time. With permission from [61].*

Experimental confirmation of the dynamical heterogeneities in colloidal repulsive glasses was given by Weeks et al. [60] using confocal microscopy in a colloidal system, which allows access to the position of all particles. In attractive colloids, similar experiments have also shown the existence of dynamical heterogeneities [61, 62], revealed by studying the distribution of displacements (in Fig. 21 for one dimension), which do not obey a Gaussian distribution. Different populations of particles with different mobility can be identified, associated with the structural heterogeneities [63].

### 4.2.4. High order singularities.

As described in the theoretical section, MCT predicts a high order singularity when two glass transitions with different driving mechanisms coexist, in the vicinity of the merging point [8]. In attractive colloids, the attraction and repulsion driven glasses meet in the region of high density and strong interactions, and the particular type of singularity depends on the details of the interactions [37]. Perhaps the most impressive consequence of the presence of this high order singularity is noted in the density correlation function, where a logarithmic decay from the short time relaxation to the structural one is predicted.

This signature has been thoroughly looked for and found in the predicted region. In Fig. 22, evidence from simulations of the binary SW system is shown [64], where the logarithmic decay is clearly visible for the marked wavevectors. The curvature of the correlation function changes from concave to convex at a particular wavevector q*, where the logarithmic trace is most clear. All these features are correctly predicted by MCT and found in different simulations irrespective of the details of the attraction (AO or SW) [64,65].

Experiments have been performed on different systems where attractive interactions could be induced, such as colloid polymer-mixtures [39,66] or micellar systems [67]. Fig. 23 shows the correlation functions of these systems obtained from dynamic light scattering, confirming the existence of the high order singularity in the region where MCT predicts it.

### 4.2.5. Attractive glasses at low density. Gelation.

In the previous sections, we have shown the attraction driven glass transition that is present at high density, where MCT correctly predicts many of its properties, even at a quantitative level. In this section we discuss how this transition extends to lower density and eventually connects to the experiments of colloidal gelation and with the well know aggregation regime DLCA (Diffusion Limited Cluster Aggregation) [41]. As the density is lowered, fewer particles are available to form the percolating structure, resulting in thinner or less branched gels. Therefore, stronger attractions are necessary to provoke dynamic arrest, i.e. the glass line moves to higher attraction strengths, or lower temperatures. Accordingly, MCT predicts for these transitions that the localization length decreases with decreasing density.

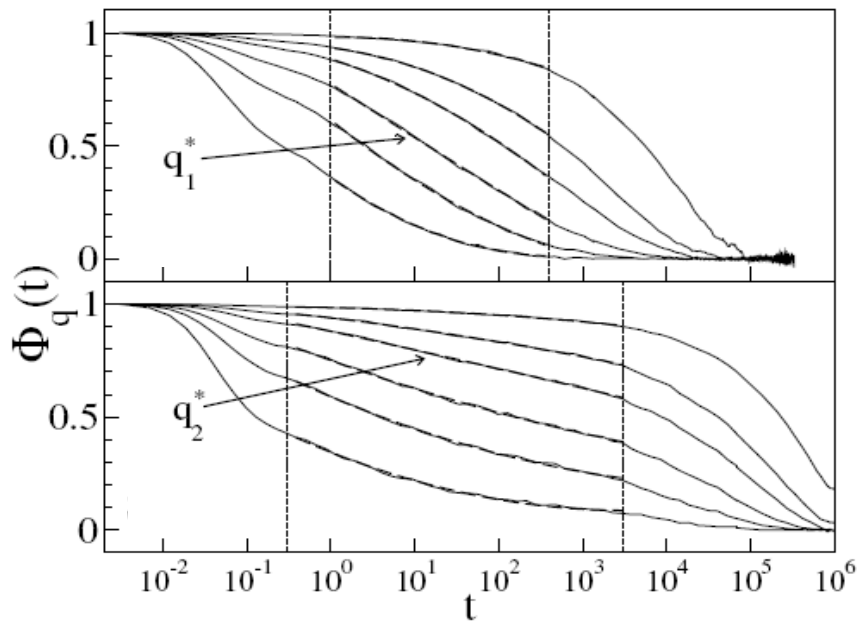

**Figure 22:** *Density correlation function for two SW systems: 3.1% width (upper panel) and 4.3% width (lower panel), both with the same volume fraction and temperature: φ = 0.6075 and T=0.4. The former has an A3 type singularity, and the latter an A4-type one. Different lines correspond to different wavevectors and dashed lines are fits to a second order expansion in log(t) in the range marked by the vertical lines. With permission from [64].*

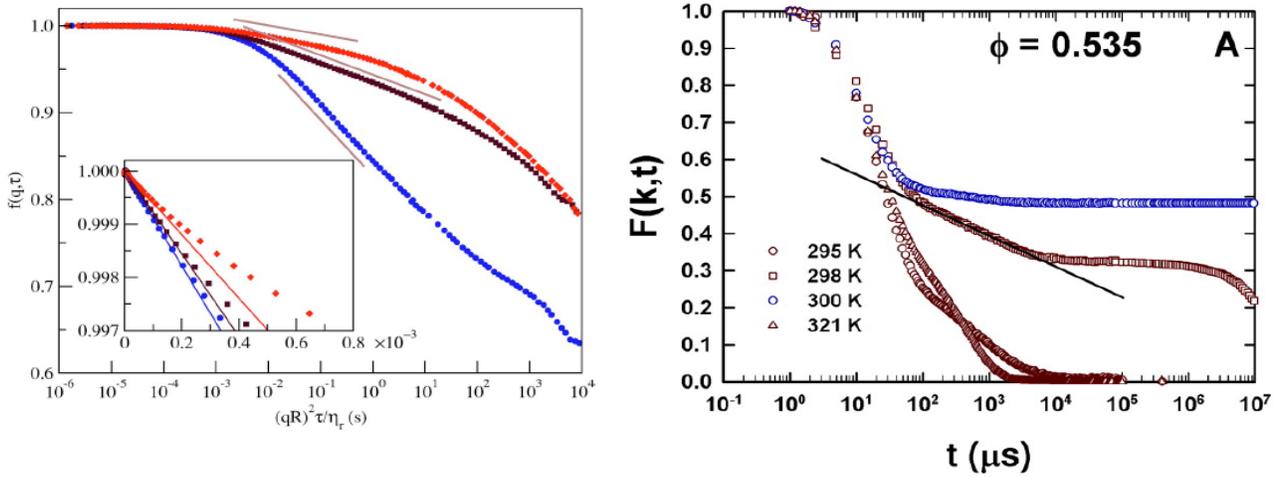

**Figure 23**: *Experimental evidences of the logarithmic decay of the density correlation functions. In left panel, a colloid-polymer mixture with constant colloid volume fraction $\phi \approx 0.64$ the blue curve is a repulsive glass, the dark one is a fluid state between the two glasses, and the red one is an attractive glass [39] – with permission from [39]. In the right panel micellar systems where the attraction is induced by increasing temperature) [67] – with permission from [67].*

However, because the density of the gel is lower, it is more flexible and flapping motions of branches or arms cause decorrelation at larger distances for lower density, even though the whole structure may not relax. This is indeed observed in simulations of attractive systems, as shown in Fig. 24 for the ideal models presented above, the maximum valence, $n_{max}$, (left panel) and the AO system with long range repulsion (right panel). The results for the $n_{max}$ model show that the plateau in the mean squared displacement increases with decreasing density (at constant temperature T=0.125; note that the transition in this model is at T=0, as shown in Fig. 19), signaling less tight structures (the localization length in the inset shows this trend). Accordingly, the non-ergodicity parameter decreases at constant wavevector, as shown in Ref. [54] for the $n_{max}$ model, and in the right panel for the AO system with repulsive barrier. It is interesting to note that in the AO system, upon decreasing the density from $\phi = 0.55$ the non-ergodicity parameter first increases (following the trend in MCT), and then decreases, whereas in the $n_{max}$ model it decreases continuously, in agreement with the different mechanisms in both systems, that ultimately causes the transition line at zero temperature in the latter. With competing interactions, it has been shown that the fluid-fluid transition can be avoided also at low density, but ordered compact local structures form, namely the Bernal spiral, which are capable of provoking dynamic arrest, even with a small number of neighbours [28,68].

Calculations with MCT have been attempted by Wu et al. [69] using a model with a short range Yukawa attraction supplemented by a long range Yukawa repulsion. The structure factors and non-ergodicity parameters reproduced the low wavevector peak, or cluster peak, which grows and exceeds the neighbour peak, i.e. compatible with arrested phase separation. However, although these calculations were extended to very low volume fractions, the trend observed in the figure above for the localization length was not reproduced, i.e. the flapping modes are missing in the theory. It is therefore necessary to incorporate them in MCT "by hand" or alternatively, develop a scheme where these modes appear naturally.

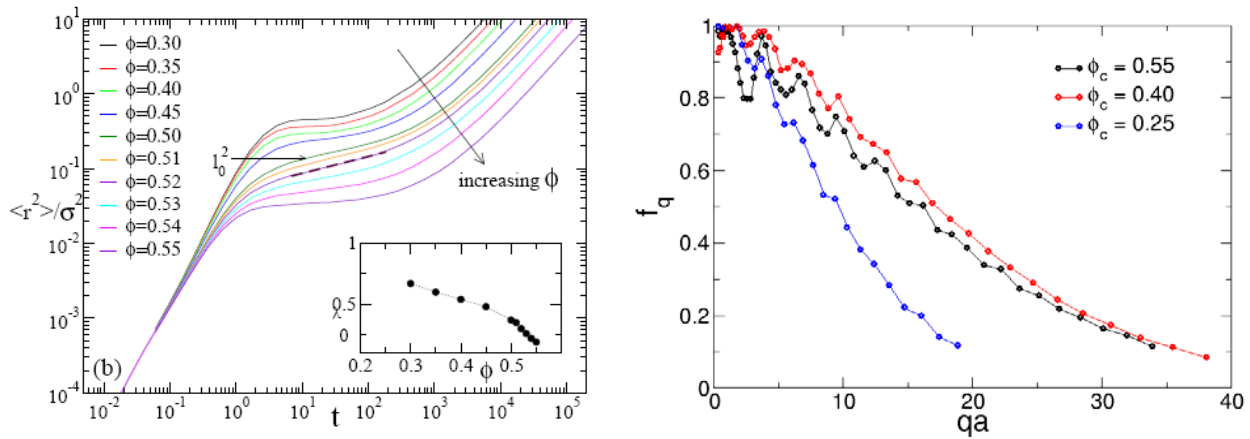

**Figure 24:** *Extension of the attractive glass to low densities. Mean squared displacement of fluid states close to the attraction driven glass (at constant temperature T=0.125) for the limited valence model ($n_{max}$=4) – left panel (taken from [54] with permission). The inset in this panel shows the evolution of the localization length with the packing fractions. Critical non-ergodicity parameters for the AO system with barrier at the densities indicated (right panel).*

These results introduce an important difference between attraction driven glasses and gels, with regard to the mechanisms for structural relaxation. This difference is also apparent in the formation of the gel. When a fluid is "quenched" beyond its glass point, the dynamics arrests progressively; the structural relaxation takes longer and longer, but the non-ergodicity parameter does not grow, i.e. structural changes are not occurring. This phenomenology has been described with computer simulations in atomic glasses or repulsion driven colloidal glasses [70], and recently in attraction driven glasses [43]. For the latter, the typical behaviour of the density correlation function is shown in Fig. 25 for the AO system with repulsive barrier (left panel).

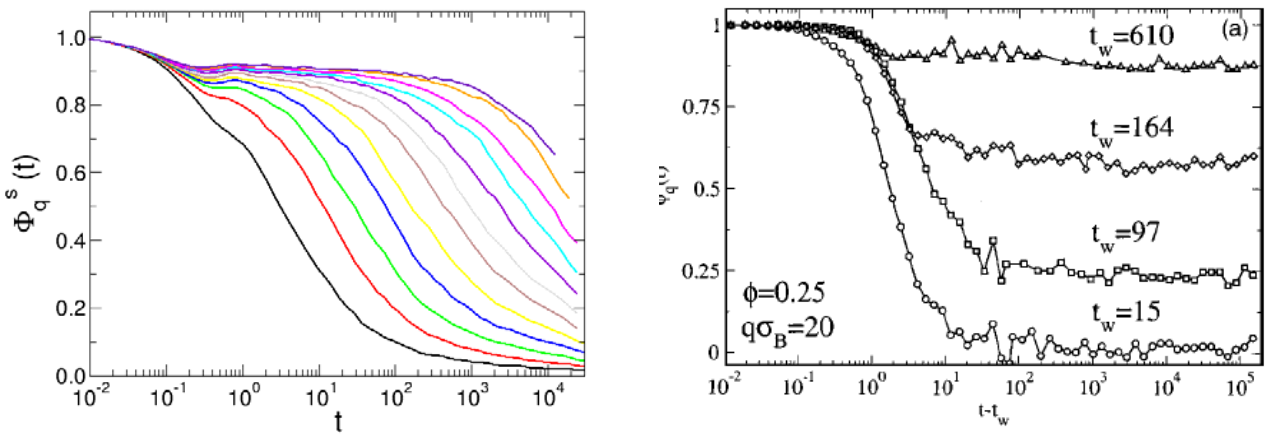

**Figure 25**: *Density correlation function for the AO system with repulsive barrier at $\phi = 0.40$ and $\phi_p = 0.80$ -left panel- and for the SW system at low density (liquid-gas separation stops due to the dynamic arrest) –right panel-. Note that the plateau increases with waiting time after the quench, similar to experimental gels, due to the formation of denser and more arrested structures. Right panel reproduced with permission from [44].*

In contrast, in colloidal gels the formation of the structure competes with structural relaxations, and a first process where the non-ergodicity parameter increases with the waiting time since the quench is found, followed by the typical aging of glasses (increase of structural relaxation time with the age of the sample). The right panel in the figure shows the formation of bonds in a SW mixture [44], but not the increase of the structural relaxation time. Evidence from experiment and simulations for this type of aging of colloidal gels has been provided by several groups [71, 72], but also low density gels with the typical aging of dense glasses are observed [73].

### 4.2.6. Discussion. Gels and attractive glasses.

The results presented above show the existence of a glass transition that is driven by the formation of quasi-permanent bonds between particles, due to a short range attraction between particles. This transition can occur in the region where fluid-fluid separation is favoured thermodynamically, and therefore a competition between both processes can take place. Several models have been designed to avoid this competition, and to isolate the effects of the glass transition.

In experiments, however, one cannot prevent phase separation and the final stage of the system is controlled by the competition. It is now generally accepted that most of the gels found experimentally are the result of arrested demixing, due to the crossing of the attractive glass line with the liquid branch of the spinodal [74, 75]. In this picture, liquid-gas separation would be obtained in the temperature range between the critical point and the crossing point, and gels for lower temperatures (stronger attractions). Only when the attractive glass line is crossed, phase separation is arrested even for short times, and more homogeneous systems are obtained [76]. It is also argued that for short range interactions, the fluid-fluid binodal is very flat, and the glass line follows it; therefore any quench inside the phase separation region results in a gel, according to the absence of liquid-gas separation in many experimental systems [77]. In any case, it is now evident that a process that produces density fluctuations and stabilizes dense regions in the system is necessary in order to have an attractive glass and/or gel transitions, either fluid-fluid separation or short range attractions with longer range repulsive interactions.

## 5. CONCLUSIONS

In this chapter we have presented the analysis of gelation in systems with short range attractions, mainly from the point of view of computer simulations and theory. The analysis presented here indicates that gelation at high density should be viewed as an attraction driven glass transition, in agreement with the theoretical predictions from MCT. The fluid-fluid separation, which may take place in the same region of the phase diagram, is completely arrested, and the dynamics of the system is controlled by the glassy dynamics. At low density, new relaxation mechanisms play a dominant role, which are absent at higher density, needing further theoretical developments. For the rationalization of experiments on gels, the competition between vitrification and fluid-fluid separation must be considered, especially at low density.

Within MCT, the fluid structure factor controls the equations of motion of the structural relaxation, and glass transitions are identified as bifurcation points. For fluids with short range attractions, MCT predicts an attraction driven glass transition induced by the formation of quasi-permanent bonds between particles, in addition to the repulsion driven one, caused by the caging of hard particles. In the region where these two glass transitions merge, a high order singularity is predicted, with peculiar properties of the relaxation mechanism. Experiments and simulations have shown that these predictions are fulfilled, even nearly at a quantitative level when the structure factor from simulations is taken as input for MCT. Yet, some quantitative discrepancies remain, as usually found in MCT analysis of the glass transition: the transition point is incorrectly located and different exponents are found for the divergence of the time scale and diffusion coefficient.

At the particle level, the system is heterogeneous, both structurally and dynamically. Structural heterogeneities are caused by the local clustering induced by the bonds, that open up holes in the elsewhere system – different mobilities are thus observed for particles in different environments. These dynamical heterogeneities are more dramatic in attraction driven glasses than in repulsion driven ones, but are absent in MCT, what poses an interesting dichotomy.

Interestingly, for low densities, a new mechanism for the relaxation of meso-structures (the size of a few particles) appears, i.e. cluster motion or flapping of branches. This provokes an increase of the localization length, and a concomitant decrease of the non-ergodicity parameter. It must be also noticed that theoretical calculations using MCT miss this relaxation modes. This relaxation mechanism can serve to distinguish gels from attractive glasses, as observed in dilute systems in contrast to the dense ones, with implications for the elasticity of the final structure.

Finally, several routes to gelation have been presented: systems with arrested spinodal decomposition and without states of fluid-fluid coexistence; systems where the fluid-fluid separation is arrested only at very low temperatures (but can be observed at higher ones); and systems where the fluid-fluid transition is suppressed. Whereas the former probably applies to most experiments with short range attractions (colloid-polymer mixtures), the second mechanism is found when longer ranged attractions are present (particularly in the case of proteins). The latter has been extensively studied by computer simulations since it allows access to the attractive glass transition from homogeneous fluid states, and is the main objective of the present work.

**ACKNOWLEDGEMENTS**

Financial support from the DAAD and the Spanish Ministerio de Educación y Ciencia, project HA2004-0022, the IFPRI (MF) and the MEC –project MAT2006-13646-C03-02- (AMP), is acknowledged. We thank a long list of colleagues for scientific collaboration and for very pleasant discussions, particularly Mike Cates, Johan Bergenholtz, Francesco Sciortino and Wilson Poon.